\documentclass[sigconf,screen]{acmart} 
\pdfoutput=1

\usepackage[normalem]{ulem}
\usepackage{wasysym}
\usepackage{xcolor}
\usepackage{subfig}
\usepackage{microtype}
\usepackage{siunitx}
\usepackage{xurl}
\usepackage{graphicx}

\setcopyright{acmlicensed}
\acmPrice{15.00}
\acmDOI{10.1145/3582016.3582065}
\acmYear{2023}
\copyrightyear{2023}
\acmSubmissionID{asplosc23main-p731-p}
\acmISBN{978-1-4503-9918-0/23/03}
\acmConference[ASPLOS '23]{Proceedings of the 28th ACM International Conference on Architectural Support for Programming Languages and Operating Systems, Volume 3}{March 25--29, 2023}{Vancouver, BC, Canada}
\acmBooktitle{Proceedings of the 28th ACM International Conference on Architectural Support for Programming Languages and Operating Systems, Volume 3 (ASPLOS '23), March 25--29, 2023, Vancouver, BC, Canada}
\received{2022-10-20}
\received[accepted]{2023-01-19}
\begin{document}
\title[Characterizing and Optimizing End-to-End Systems for Private Inference]{Characterizing and Optimizing End-to-End\\Systems for Private Inference}

\author{Karthik Garimella}
\email{kg2383@nyu.edu}
\affiliation{%
  \institution{New York University}
  \city{New York}
  \state{New York}
  \country{USA}
}

\author{Zahra Ghodsi}
\email{zahra@purdue.edu}
\affiliation{%
  \institution{Purdue University}
  \city{West Lafayette}
  \state{Indiana}
  \country{USA}
}

\author{Nandan Kumar Jha}
\email{nj2049@nyu.edu}
\affiliation{%
  \institution{New York University}
  \city{New York}
  \state{New York}
  \country{USA}
}

\author{Siddharth Garg}
\email{sg175@nyu.edu}
\affiliation{%
  \institution{New York University}
  \city{New York}
  \state{New York}
  \country{USA}
}

\author{Brandon Reagen}
\email{bjr5@nyu.edu}
\affiliation{%
  \institution{New York University}
  \city{New York}
  \state{New York}
  \country{USA}
}

\date{}

\setcounter{page}{1}

\thispagestyle{empty}

\begin{abstract}

In two-party machine learning prediction services, the client’s goal is to query a remote server’s trained machine learning model to perform neural network inference in some application domain. However, sensitive information can be obtained during this process by either the client or the server, leading to potential collection, unauthorized secondary use, and inappropriate access to personal information. These security concerns have given rise to Private Inference (PI), in which both the client's personal data and the server's trained model are kept confidential. State-of-the-art PI protocols consist of a pre-processing or offline phase and an online phase
that combine several cryptographic primitives: Homomorphic Encryption (HE), Secret Sharing
(SS), Garbled Circuits (GC), and Oblivious Transfer (OT). Despite the need and recent performance improvements, PI remains largely arcane today and is too slow for practical use. 

This paper addresses PI’s shortcomings with a detailed characterization of a standard high-performance protocol to build foundational knowledge and intuition in the systems community. Our characterization pinpoints all sources of inefficiency -- compute, communication, and storage. In contrast to prior work, we consider inference request arrival rates rather than studying individual inferences in isolation and we find that the pre-processing phase cannot be ignored and is often incurred online as there is insufficient downtime to hide pre-compute latency. Finally, we leverage insights from our characterization and propose three optimizations to address the storage (Client-Garbler),  computation (layer-parallel HE), and communication (wireless slot allocation) overheads. Compared to the state-of-the-art PI protocol, these optimizations provide a total PI speedup of 1.8 $\times$ with the ability to sustain inference requests up to a 2.24 $\times$ greater rate. Looking ahead, we conclude our paper with an analysis of future research innovations and their effects and improvements on PI latency.

\end{abstract}


\maketitle
\section{Introduction}

Privacy continues to increase in significance as users steadily demand more protection and control over how and when their data is used.
Addressing these concerns will have a profound effect on how many popular cloud services are implemented, which rely heavily on precise user data for high-quality experiences.
One solution is to leverage privacy-preserving computation 
to guarantee the privacy and security of user data \emph{during} computation, effectively extending the guarantees of today's methods that only protect communication (and possibly storage) to the entire execution pipeline.
Privacy-preserving computation provides the best of both worlds for the user: privacy and security are significantly improved while retaining the ability to use online services now integral to everyday life.
\begin{figure*}[!t]
\centering
\includegraphics[width=\textwidth]{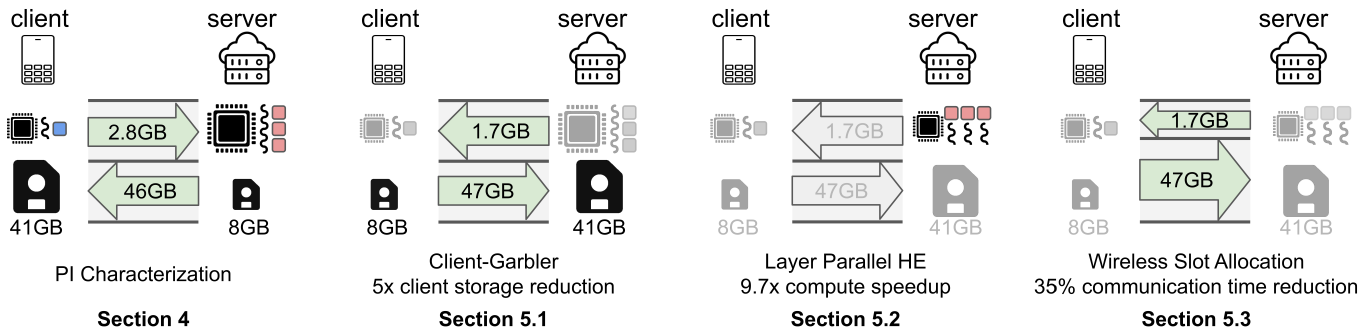}
\caption{Overview of the paper's organization and contributions in reducing overheads of hybrid PI protocols. Each image highlights a contribution and summarizes the key insight or result.}
\vspace{-.5em}
\label{fig:summary}
\end{figure*}

AI is a natural starting point for privacy-preserving computing. 
These workloads make heavy use of the client-cloud model and require access to user data to be successful.
Moreover, AI accounts for a significant amount of all private user data processing.
E.g., Meta processes over 200 trillion inferences per day~\cite{lee2019accelerating}. The ability to process certain essential functions in private (such as convolutions, ReLU, and fully-connected layers) could safeguard a disproportionate amount of users' data and would mark a significant achievement in privacy-preserving computation.
Privacy-preserving computation is still a developing field, and as we will show inference is still a ways off from being practical.
We focus on private inference (PI) and leave training to future work.

The solution space of privacy-preserving techniques is vast, and each has its strengths and weaknesses.
One way to classify techniques is in how they achieve security.
Whether by:
system implementation (e.g., SGX \cite{intel_sgx}),
statistics (e.g., differential privacy \cite{dwork2006differential,abadi2016deep}), 
or cryptography (e.g., homomorphic encryption \cite{rivest1978data,gentry2009fully,gentry2012fully,fan2012somewhat,gentry2013homomorphic,brakerski2014leveled,cheon2017homomorphic} and multi-party computation \cite{shamir1979share,even1985randomized,beaver1991efficient,beaver1995precomputing,goldreich2019play,rabin2005exchange,ishai2003extending}).
Systems solutions offer the best performance but are also less secure because
they assume hardware trust and are vulnerable to attack~\cite{van2018foreshadow}.
Statistical privacy is stronger and provides quantifiable guarantees, e.g., $\epsilon$-DP \cite{dwork2014algorithmic}.
However, these methods add noise to the computation,
which can result in degraded accuracy, and DP has been most successful in answering questions in aggregate.
Their use in computing on individual user data, e.g., a single inference, is unclear.
Cryptography-based solutions compute exact inferences with strong security guarantees (e.g., 128-bit) but come at the expense of high computation, communication, and storage overhead.
Homomorphic encryption (HE) and secure multi-party computation (MPC), including secret sharing (SS) and garbled circuits (GCs), are the two leading methods for cryptographically-secure computation and the focus of this paper.

There now exists a large body of research on private inference using both HE and MPC. At a high-level, many protocols share a common, hybrid approach that tailors cryptographic primitive selection to the computational needs of each network layer.
Most leverage homomorphic encryption and/or secret sharing for arithmetic circuits (i.e., convolutions and fully-connected layers) and multi-party computation for boolean circuits such as ReLUs~\cite{mishra2020delphi,jha2021deepreduce,lou2020safenet,cho2022selective,cho2021sphynx}.
(Non-hybrid approaches have been proposed but typically sacrifice accuracy, see Section~\ref{sec:RelatedWork}.)
Recent work has sought to further improve hybrid protocols with better neural architecture design~\cite{ghodsi2020cryptonas,cho2021sphynx}, protocol optimization~\cite{mishra2020delphi,lou2020safenet,jha2021deepreduce,cho2022selective}, and hardware acceleration~\cite{reagen2021cheetah,samardzic2021f1,samardzic2022craterlake}. 
These optimizations have largely focused on reducing specific components of the overall PI protocol, and it is unclear how they combine to improve the overall goal of fast end-to-end private inference.



This paper provides the first characterization of end-to-end private inference using state-of-the-art hybrid protocols. 
The characterization encompasses both the online and offline phases, accounting for compute, communication, and storage overheads. 
We are also the first to consider inference request arrival rates (workloads of inferences) in PI;
all known prior work focuses on individual inferences in isolation.
We find that this is a vital distinction as the offline costs are so large that they do not always remain offline, even at low arrival rates.

Our characterization provides the following  insights:
\begin{itemize}
    \item The computation time of the offline phase cannot be ignored when processing PI requests.
    It involves (HE and GC) computations that run on the order of minutes.
    
    \item Hybrid protocols incur large offline storage costs (e.g., tens of GBs).
    These must be stored on the client device, limiting the number of PI pre-computations that can be buffered and quickly overwhelm storage constraints.
    
    \item ReLUs introduce both large communication and computation overheads. For example,
    ReLU transmission and computation times for TinyImageNet using ResNet-18 can take up to 11 and 21 minutes, respectively.
\end{itemize}

Leveraging our insights,
we propose three optimizations to improve inference latency that address each of the system limitations: compute, communication, and storage.
An overview is shown in Figure~\ref{fig:summary}.
First, to overcome the limits of client storage, we propose a protocol optimization that we refer to as the \textbf{Client-Garbler} protocol.
This reverses the client and cloud roles in GC, enabling pre-computed GCs to be stored on the server, which we assume has significantly more storage than a client's smartphone.
Second, we identify a form of compute parallelism termed \textbf{layer-parallel HE} (LPHE). 
The observation is that each neural network layer's offline HE computation is independent
and can be run completely in parallel. 
LPHE significantly reduces the offline computational costs of hybrid protocols, reducing their impact on inference latency even when they are incurred online. Furthermore, LPHE implies that hybrid solutions using a combination of protocols (i.e., SS+HE) for linear layers can be faster than single-primitive solutions (i.e., HE) 
even though hybrid protocols entail more total computation.
This is because LPHE provides an opportunity to run all offline HE layer computations in an embarrassingly parallel fashion.
Finally, we show how the default provisioning of wireless bandwidth between upload and download is sub-optimal for PI and propose \textbf{wireless slot allocation} (WSA).
WSA reduces communication time by optimally dividing wireless bandwidth between upload and download based on the protocol's needs. 
Combined, the proposed optimizations improve the mean inference latency by $1.8\times$ over the state of the art~\cite{mishra2020delphi}.

This paper makes the following contributions:
\begin{enumerate}
    \item The first end-to-end characterization of private inference using arrival rates. Our analysis reveals the key sources of inefficiency with respect to storage, communication, and computation.
    
    \item A protocol optimization (Client-Garbler) to reduce the significant storage pressure on the client by $5\times$.
    
    \item Identifying a new form of parallelism (LPHE) that enables each HE linear layer computation to run simultaneously, significantly improving the performance of one of the slowest legs of PI by $9.7\times$. LPHE is favorable  when compared to Request-Level Parallelism when client storage is limited (e.g., \SI{32}{GB} and less).
    
    \item Wireless bandwidth slot allocation (WSA) to reduce communication latency by up to $35\%$.
\end{enumerate}

We conclude with a discussion of our results and their implications in Section~\ref{sec:discussion}, including the tradeoffs of different PI protocols with respect to systems and by estimating PI performance as advances to each part of the protocol are inevitably made, e.g., from HE and GC accelerators.
Our discussion points to open problems that require further research attention.
\section{Cryptographic Primitives and PI}
\label{sec:background}

This section provides a primer on the cryptographic primitives used to achieve private inference: homomorphic encryption (HE), secret sharing (SS), garbled circuits (GC), and oblivious transfer (OT).
We provide technical details and highlight their system implications.
We then outline the entire process taken to achieve high-performance private inference, including all phases of execution.
The threat model assumed is consistent with other work: semi-honest, two-party computation with no hardware and software trust.

\subsection{Cryptographic Primitives}

\subsubsection{Homomorphic Encryption}
HE is a type of encryption where encrypted data is malleable. 
This enables functions to be computed directly on ciphertexts, preserving data confidentiality during computation.

\textit{Details:} HE involves three steps. First, the client encrypts data $x$ using an encryption function $E$ and its public key $k_{pub}$, i.e., $c = E(x,k_{pub})$ (or just $c = E(x)$ for short).
Then, given two ciphertexts 
$c_1 = E(x_1)$ and $c_2 = E(x_2)$, the server homomorphically computes $x_1 \odot x_2$ using a function $\star$ that operates directly on ciphertexts, i.e., $c'=c_1 \star c_2$.  The client can then decrypt $c'$ using a decryption function $D$ and secret key $k_{sec}$ to obtain $x_1 \odot x_2 = D(c',k_{sec})$ (or just $D(c')$ for short). 

\textit{Systems implications:} HE introduces a large computational overhead of 4-6 orders of magnitude slowdown~\cite{reagen2021cheetah,samardzic2021f1,samardzic2022craterlake}. However, its storage and communication costs are minimal.

\subsubsection{Additive Secret Sharing}
Additive secret sharing (SS) allows a value $x$ to be shared amongst two (or more) parties such that neither party learns anything about $x$ from its share.
The parties can compute arithmetic functions of secret shared values to obtain shares of the result. These can then be combined to reveal the result.

\textit{Details:} 
The shares of a value $x$ are 
$\langle x\rangle_1 = r$ and $\langle x\rangle_2 = x-r$, where $r$ is a random value and the arithmetic operations are modulo a prime $p$. 
Since $x=\langle x\rangle_1+\langle x\rangle_2$, the value $x$ can be recovered if the two parties add their shares together. However, since $r$ is random, neither party has enough information to reconstruct $x$ by itself.
Given shares of two values $x$ and $y$, shares of $x+y$ can be computed by as follows:  $\langle x + y\rangle_1 =  \langle x \rangle_1 + \langle y\rangle_1$, and $\langle x + y\rangle_2 =  \langle x \rangle_2 + \langle y\rangle_2$. That is, the shares of a sum of two values equal the sum of the shares. Multiplying two secret shared values is more challenging, however. One solution is to generate so-called ``Beaver triples" in a pre-processing phase. Beaver triples are secret shares of values $a$, $b$ and $c$ where $a$ and $b$ are random and unknown to the parties, and $c= a \odot b$. Beaver triples can be generated using offline HE and leveraged online for fast private multiplications; see \cite{mishra2020delphi} for more details. 

\textit{Systems implication:} Additions in SS add minimal storage and communication overheads compared to plaintext computation. However, multiplications are costlier and require expensive HE computations in a pre-processing phase. Hybrid protocols leverage offline HE followed by online SS for linear layer computations, thus incurring large offline costs that cannot always be hidden. 

\subsubsection{Garbled Circuits}
The protocols described so far enable private arithmetic computations; additions and multiplications over integer values. GCs enable two parties to compute a \emph{Boolean} function on their private inputs without revealing their inputs to each other. Unlike SS and HE, operating over Boolean gates enables the parties to jointly compute any function. (We note that binary constructions of HE and SS exist, we discuss them in Section~\ref{sec:RelatedWork}.)

\textit{Details:} In GCs, a function is first represented as a Boolean circuit.
One party (the garbler) assigns two random labels (keys) to each input wire of each gate, labels correspond to the values $0$ and $1$. 
For each gate, the garbler then generates an encrypted truth table that maps the output labels to the gate's input labels.
The garbler sends the generated garbled circuit to the other party (the evaluator), along with the labels corresponding to its inputs. 
The evaluator then uses the OT protocol (see below) to obtain the labels corresponding to its inputs 
without the garbler learning the input values.
At this point, the evaluator can execute the circuit in a gate-by-gate fashion, without learning intermediate values. 
Finally, the evaluator shares the output labels with the garbler who maps them to plaintext values.

\textit{System implications:} GCs, like SS, involve both the client and the server in the actual computation; one must be the garbler and the other evaluator. 
The process of evaluating and garbling gates is compute-intensive, involving up to two AES computations (and additional pre-/post-processing logic) per truth table entry.
Recent optimizations have been made to improve GC compute (i.e., FreeXOR~\cite{kolesnikov2008improved} and HalfGate~\cite{zahur2015two}); we employ these optimizations and note that run-times are still long.
In addition, the wires and tables must be transmitted between the parties and stored by the evaluator prior to being used. As we will show, this introduces substantial communication and storage overheads on the system.

\begin{figure*}
\centering
\includegraphics[width=0.99\textwidth]{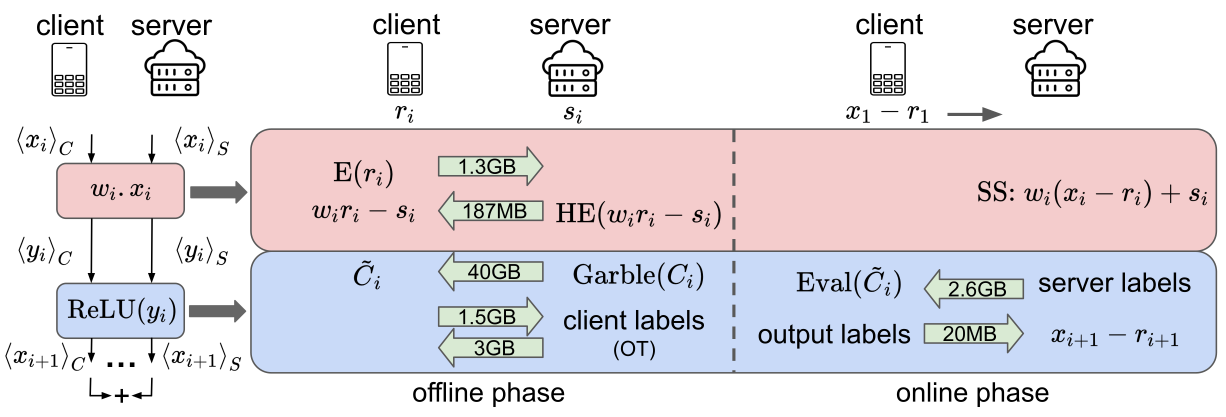}
\caption{Baseline PI protocol using HE and SS for linear layers and GC for ReLU layers for ResNet-18 on TinyImageNet.}
\label{fig:pi_protocol}
\end{figure*}

\subsubsection{Oblivious Transfer}
OT~\cite{rabin2005exchange} enables a receiver to learn only one of two values sent by a transmitter without the transmitter learning which value was learned. It is a fundamental building block in MPC and the foundation for many constructions including GCs. 

\textit{Details:}
In 1-out-2 OT~\cite{rabin2005exchange} a sender has two input values $x_0, x_1$, and the receiver has a selection bit $r$. The OT protocol allows the receiver to obtain $x_r$ without learning $x_{1-r}$ and without the sender learning $r$. The simplest OT protocol requires expensive public-key cryptography, but can be modified to use cheaper symmetric key cryptography by \emph{extending} a small number of \emph{base OTs}~\cite{ishai2003extending}.  

\textit{System implications:} The computational and storage costs of OT are relatively small, especially with OT extensions. OTs do add moderate communication overheads, but even these are small compared to the communication costs of transmitting garbled circuits. Some of our protocol optimizations add extra OT costs, but given that OTs themselves are relatively cheap, the overall impact on our optimizations is small as well.

\subsection{Private Inference} \label{subsec:PrivateInference}

We begin  by reviewing  a class of hybrid PI protocols that
use HE and SS
for linear layers, and GCs for ReLU layers~\cite{liu2017oblivious, mishra2020delphi}. These protocols have achieved state-of-the-art results in the 2PC setting.
In this setting, we assume that a client wishes to obtain an inference on her private data using the server's proprietary deep learning model. The server should not learn the client's input and the client should not learn the parameters of the server's model.
Following prior work, we assume both parties are honest-but-curious, i.e., they follow the protocol faithfully but try to extract information during protocol execution.
We protect against the same threat models and refer to prior work, e.g., DELPHI~\cite{mishra2020delphi}, for details.

We briefly describe DELPHI~\cite{mishra2020delphi}, one such hybrid protocol (see Figure~\ref{fig:pi_protocol}) that we assume as a baseline in this paper.
The protocol consists of an offline phase, which only depends on the network architecture and parameters, and an online phase, which is performed after the client's input is available.
(One of the major insights in DELPHI was to move many of the expensive computations to the offline phase to improve online inference latency.)
We represent model parameters at layer $i$ with $\mathbf{w_i}$ (omitting biases in the description for simplicity). The layers after each linear and ReLU operation are represented by $\mathbf{y_i}$ and $\mathbf{x_i}$, respectively.
At a high level, the protocol allows the client and the server to hold additive shares of each layer value during inference evaluation.

\textbf{Offline Phase.}
During the offline phase, the client generates the encryption keys and sends them to the server. The client and the server then sample random vectors per each linear layer
from a finite ring,
denoted by $\mathbf{r_i}$ and $\mathbf{s_i}$ respectively. The client homomorphically encrypts its random vectors and sends them to the server. After obtaining $\textrm{E}(\mathbf{r_i})$, the server computes $\textrm{E}(\mathbf{w_i}.\mathbf{r_i}-\mathbf{s_i})$ homomorphically and sends it back to the client. The client decrypts the ciphertext received from the server and stores the plaintext values.
For each non-linear ReLU operation, the server constructs and garbles a circuit that implements $\textrm{ReLU}(\mathbf{y_i})-\mathbf{r_{i+1}}$, where $\mathbf{r_{i+1}}$ is the next layer randomness (provided by the client).
The server sends the garbled circuits to the client and the two parties engage in OT after which the client obtains labels corresponding to its inputs.
At the end of the offline phase, the server stores its random vectors $\mathbf{s_i}$ and the garbled circuit input and output encodings, and the client stores its random vectors $\mathbf{r_i}$, garbled circuits, and the labels corresponding to its inputs. The communication and storage requirements for ResNet-18 inference on a single input from TinyImageNet are shown in Figure~\ref{fig:pi_protocol}.

\textbf{Online Phase.} 
The client's input $\mathbf{x_1}$ is available in the online phase. The client computes $\mathbf{x_1}-\mathbf{r_1}$ and sends it to the server. At the beginning of $i$th linear layer, the server holds
$\mathbf{x_i}-\mathbf{r_i}$ and the client holds $\mathbf{r_i}$.
The server computes its share of layer output as $\langle \mathbf{y_i} \rangle_{s} = \mathbf{w_i}(\mathbf{x_i}-\mathbf{r_i})+\mathbf{s_i}$. 
The client holds $\langle \mathbf{y_i} \rangle_{c} = \mathbf{w_i}.\mathbf{r_i}-\mathbf{s_i}$ from the offline phase, and the client and the server hold additive shares of $\mathbf{y_i}=\mathbf{w_i}.\mathbf{x_i}$. To evaluate the ReLU layer, the server obtains the labels corresponding to its share of ReLU input $\langle \mathbf{y_i} \rangle_{s}$ and sends them to the client. The client now holds labels for the server's input, as well as labels for its inputs $\langle \mathbf{y_i} \rangle_{c}$ and $\mathbf{r_{i+1}}$, which were obtained during the offline phase along with the garbled circuits. The client evaluates the circuit and sends the output labels to the server who is then able to obtain
$(\mathbf{x_{i+1}}-\mathbf{r_{i+1}})$. At this point, the client and server hold additive shares of $\mathbf{x_{i+1}}$ and will similarly evaluate subsequent layers.

\textbf{Hybrid versus HE-only PI.} In contrast to hybrid protocols, 
HE-only PI protocols do not have an offline phase; all PI computation and communication takes place once the client’s input is available \cite{gilad2016cryptonets}. 
Specifically, the client encrypts their input $\mathbf{x_1}$ and sends it to the server to process the inference under HE.
The server then sends the encrypted prediction to the client who can decrypt the ciphertext using their private key.
However, in HE-only protocols, ReLU (and other non-linearities) must be replaced by polynomial functions in order to be supported by HE operations.
This use of polynomials has been shown to greatly reduce network accuracy, especially for deeper networks \cite{garimella2021sisyphus}.
The hybrid PI protocols used in this paper preserve accuracy by implementing ReLU using garbled circuits rather than using polynomial approximations.

Given the high cost of performing HE computations during the online phase, recent HE accelerators have been proposed to reduce compute latency ~\cite{reagen2021cheetah, samardzic2021f1, samardzic2022craterlake}.
These HE accelerators are evaluated with shallow networks (6, 10, and 20 layers) and small-input datasets such as MNIST and CIFAR-10.
Cheetah~\cite{reagen2021cheetah} looks at deeper networks (e.g., ResNet50), and follows Gazelle's~\cite{juvekar2018gazelle} hybrid protocol.
However, Cheetah focuses on only accelerating the convolution and fully connected layers using HE and does not consider the processing time of ReLU, see Section 2A in paper~\cite{reagen2021cheetah}. For example while Cheetah reports an HE processing time of 198ms  for ResNet-50 on ImageNet (Table 6 in~\cite{reagen2021cheetah}), the total end-to-end inference latency would be 215 seconds when considering both the online computation and communication required for evaluating the 9.6  million ReLUs in ResNet-50 (we observe \SI{83}{\SIUnitSymbolMicro s} and \SI{2.05}{KB} per ReLU which is in agreement with Table 3 from \cite{mishra2020delphi}). This calculation optimistically assumes 4 threads for ReLU evaluation as Gazelle and Delphi do as well as \SI{10}{Gbps} bandwidth (equivalent to LAN, as used in \cite{mp2ml}). 
Cheetah could be used to speed up the HE compute in this work, however, we optimize beyond accelerators such as 
Cheetah by analyzing system-level bottlenecks caused by not only HE latency, but also GC latency, communication, and storage.
As we consider and optimize the full system, the PI latency reported here is much higher;
ReLUs are expensive in PI.







\section{Methodology}
\label{sec:methodology}

We perform experiments using ResNet networks~\cite{he2016deep}
(ResNet-32 and ResNet-18) and VGG-16~\cite{vgg} on the CIFAR-100~\cite{cifar} and TinyImageNet~\cite{yao2015tiny} datasets.
As in prior work, we remove downsampling and replace max-pooling with average-pooling~\cite{jha2021deepreduce, ghodsi2021circa}). 
CIFAR-100 has images of spatial resolution 32$\times$32 while TinyImageNet's images are 64$\times$64. 
Generally, from ResNet-32, VGG-16, and to ResNet-18 the number of parameters, FLOPs, ReLUs, and accuracy increases.

We stand apart from prior work by modeling the client device after an Intel Atom Z8350 embedded device (\SI{1.92}{GHz}, 4 cores, \SI{2}{GB} RAM) and the server as an AMD EPYC 7502 (\SI{2.5}{GHz}, 32 cores, \SI{256}{GB} RAM). Prior work runs the client-side computation on server hardware.
We measure the latency, storage, and communication costs for the online and offline phase using the DELPHI codebase~\cite{mishra2020delphi}. 
For all networks, we averaged these costs over 10 and 5 runs for CIFAR-100 and TinyImageNet, respectively. We implement LPHE using DELPHI's implementation of Gazelle's~\cite{juvekar2018gazelle} HE evaluation algorithm and benchmark the HE computations for each linear layer used. These HE operations are implemented using SEAL~\cite{sealcrypto}. When using LPHE, we set the number of  threads to the number of linear layers.

We use these measurements to construct a model of a system for PI and a simulator using Simpy~\cite{simpy} to explore and evaluate tradeoffs under different system conditions. We model a single-client and single-server system where inference requests are queued and served in a FIFO manner. We generate inference requests following a Poisson distribution, as in prior work~\cite{hauswald2015sirius,li2016work,kasture2016tailbench,gan2019open,reddi2020mlperf,gupta2020deeprecsys}. 
We validate our simulator against DELPHI by configuring our simulator to match their system and measure the compute latency of GC garbling, GC evaluating, and HE linear evaluation: we report a relative error of $0.9\%$ for TinyImageNet on ResNet-18.
Our code is available at the following
GitHub repository: \href{https://github.com/kvgarimella/characterizing-private-inference}{https://github.com/kvgarimella/characterizing-private-inference}.

\begin{figure}[!t]%
    \centering
    
    {{\includegraphics[width=0.99\columnwidth]{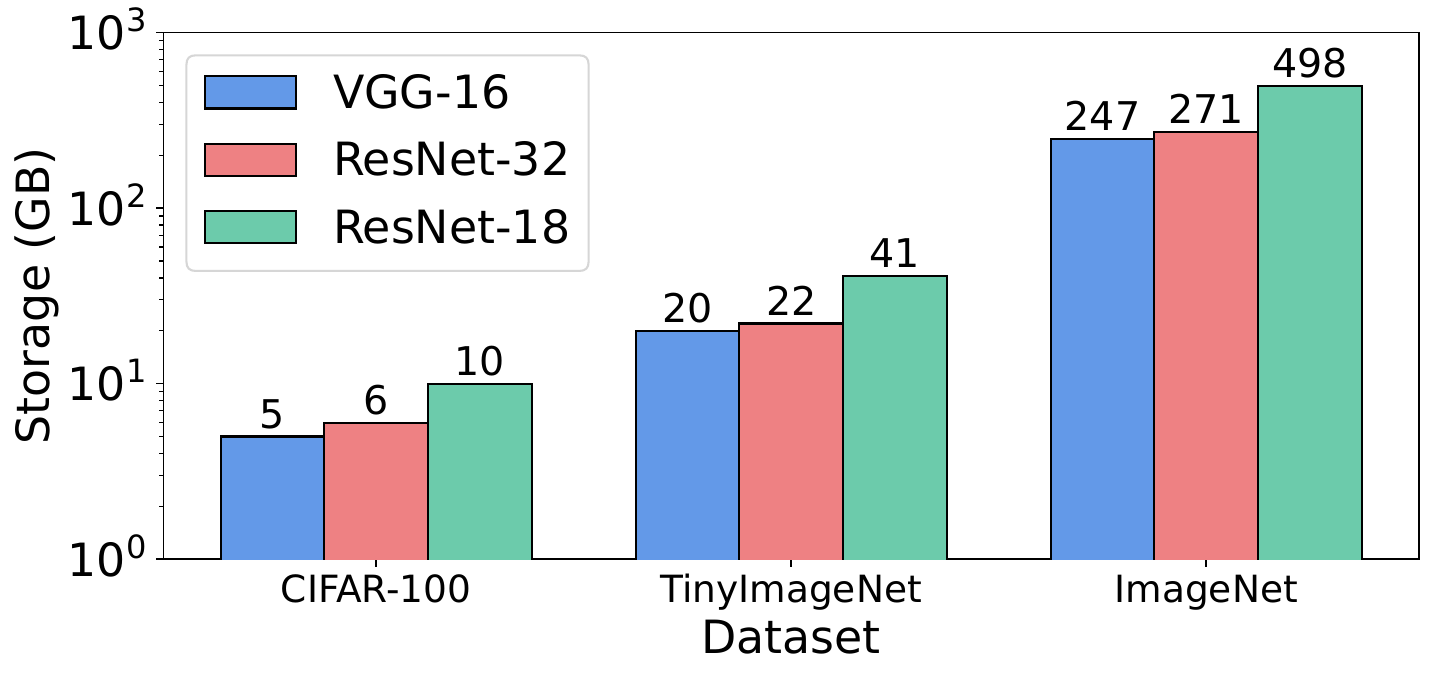} }}
    \vspace{-2em}
    \caption{Total amount of pre-processing data that must be stored on the client's device \textit{per inference}. Garbled Circuits make up a majority of this storage cost.}
    \label{fig:storage_issue}%
\end{figure}

\section{Performance Characterization of PI}
\label{sec:characterization}

This section presents our PI profiling results for multiple datasets and networks.
We start by characterizing the overheads of individual inferences and then study the impact of a stream of requests on mean inference latency. We note that all results in this section use the baseline Server-Garbler protocol.

\subsection{Single Inference Overheads}

\subsubsection{Storage:}
ReLUs dominate the storage overheads of PI. ReLU GCs are generated by the server in the offline phase and stored by the client for later use
during the online phase.
Profiling the fancy-garbling library~\cite{ball2019garbled,fancygarble} we find that the server (garbler) incurs a modest storage penalty of \SI{3.5}{KB} per ReLU to store the encoding information for inputs to the GCs.
The client (evaluator), however, incurs a $5\times$ greater storage cost of \SI{18.2}{KB} per ReLU from storing the GCs themselves. 

Figure~\ref{fig:storage_issue} shows the total client storage costs of PI 
for different datasets and networks.
For networks run on smaller inputs (CIFAR-100), 
the client must have at least 5 to \SI{10}{GB} of storage available for 
a \emph{single inference} (note that these pre-computed values cannot be reused or amortized over multiple inputs). 
Larger inputs and larger networks have more ReLUs and storage costs go up proportionally.
For TinyImageNet, which has been used extensively in PI research, the most accurate model (ResNet-18) requires  \SI{41}{GB} \emph{per inference}. This is a major obstacle for PI: The global average smartphone storage capacity is nearly 100GB \cite{average_smartphone_storage}, meaning that an individual would need to allocate nearly half of their total storage for just a single inference pre-compute. Prior work, however, has not contended with this limitation.

Figure~\ref{fig:storage_issue} also reports the storage overheads of ImageNet,
which are on the order of hundreds of GBs. 
Storing (and communicating) roughly half a terabyte of data per inference is beyond the capabilities of current technology; ImageNet is therefore not commonly studied in the PI literature.


\begin{figure}[!t]%
    \centering
    {{\includegraphics[width=0.99\columnwidth]{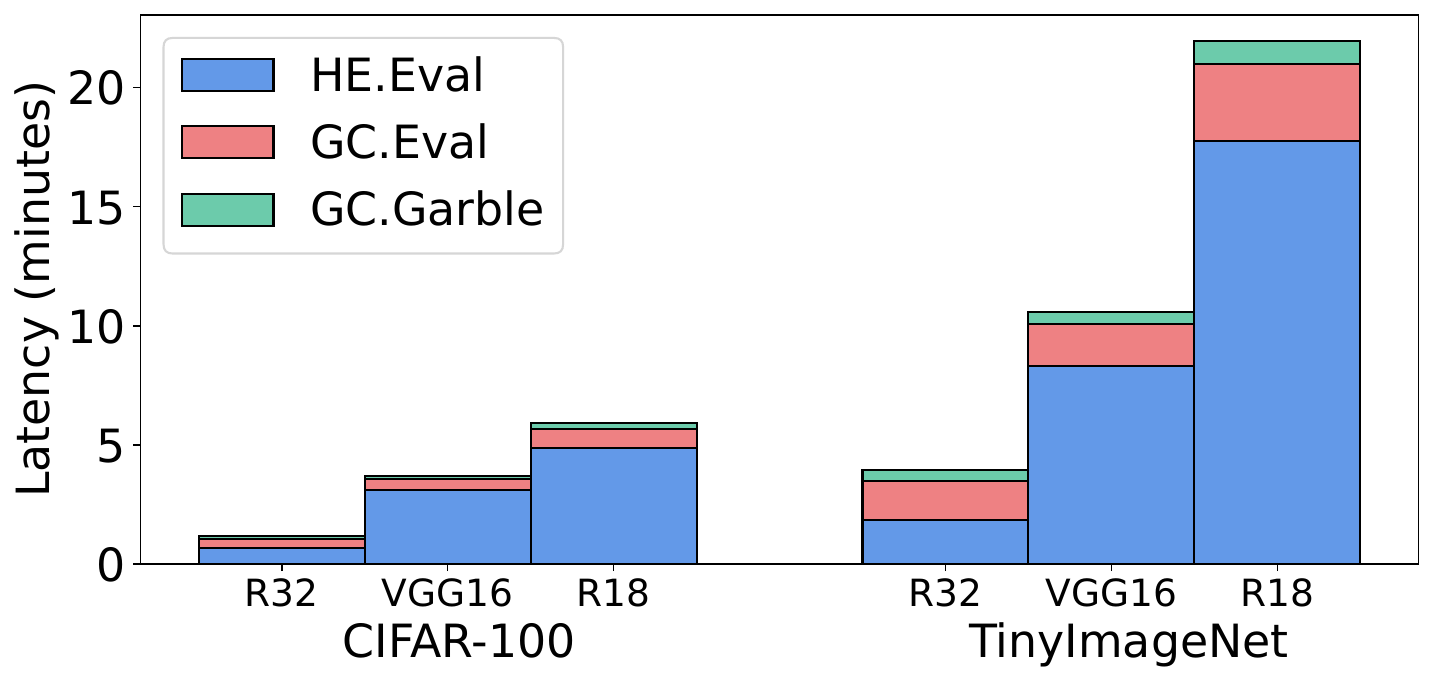} }}
    \vspace{-2em}
    \caption{Latency of homomorphic evaluations for linear layers (HE.Eval),  GC garbling (GC.Garble) and GC evaluation (GC.Eval) \textit{per inference}. GC.Eval occurs client-side and the rest are processed on the server.}%
    \label{fig:compute_issue}%
\end{figure}

\subsubsection{Computation:}
In the baseline Server-Garbler protocol, the server is responsible for executing HE computation, GC garbling, and SS evaluation while
the client must process GC evaluations. Secret shares generated via HE and GC garbling are performed in the offline phase, leaving only GC evaluation and the server's SS evaluation as the computation that must happen online.

Figure~\ref{fig:compute_issue} shows the latency breakdown of these cryptographic primitives for a single inference. 
As can be seen, offline HE computations of linear layers (blue) dominate the total compute cycles for PI protocols. 
Though smaller than HE, GC evaluations (red) are a sizeable fraction of overall computational cost (and dominate online costs), while offline GC garbling (green) cost is almost negligible. 
The asymmetry between GC evaluation and garbling is apparent because we characterize the client (evaluator) using a realistic client device that is much less powerful than the server (garbler).
These observations highlight two reasons prior work is overly optimistic in its evaluation of PI costs:
(1) by assuming that offline costs can be hidden via pre-computation, prior work ignores an otherwise dominant component of the overall cost; and (2) online costs are also underestimated if the computational asymmetry between client and server is ignored.
We do not plot SS evaluation as its run-time is negligible when compared to HE and GC (e.g., 0.61 seconds for ResNet-18 on TinyImageNet).


\subsubsection{Communication:}
Hybrid PI protocols require multiple rounds of interaction between the client and  server during both the offline and online phases, as shown in Figure~\ref{fig:pi_protocol}. The dominant communication costs are from the GC protocol, including its OT sub-protocol. 

\begin{figure}[!t]%
    \centering
    {{\includegraphics[width=0.99\columnwidth]{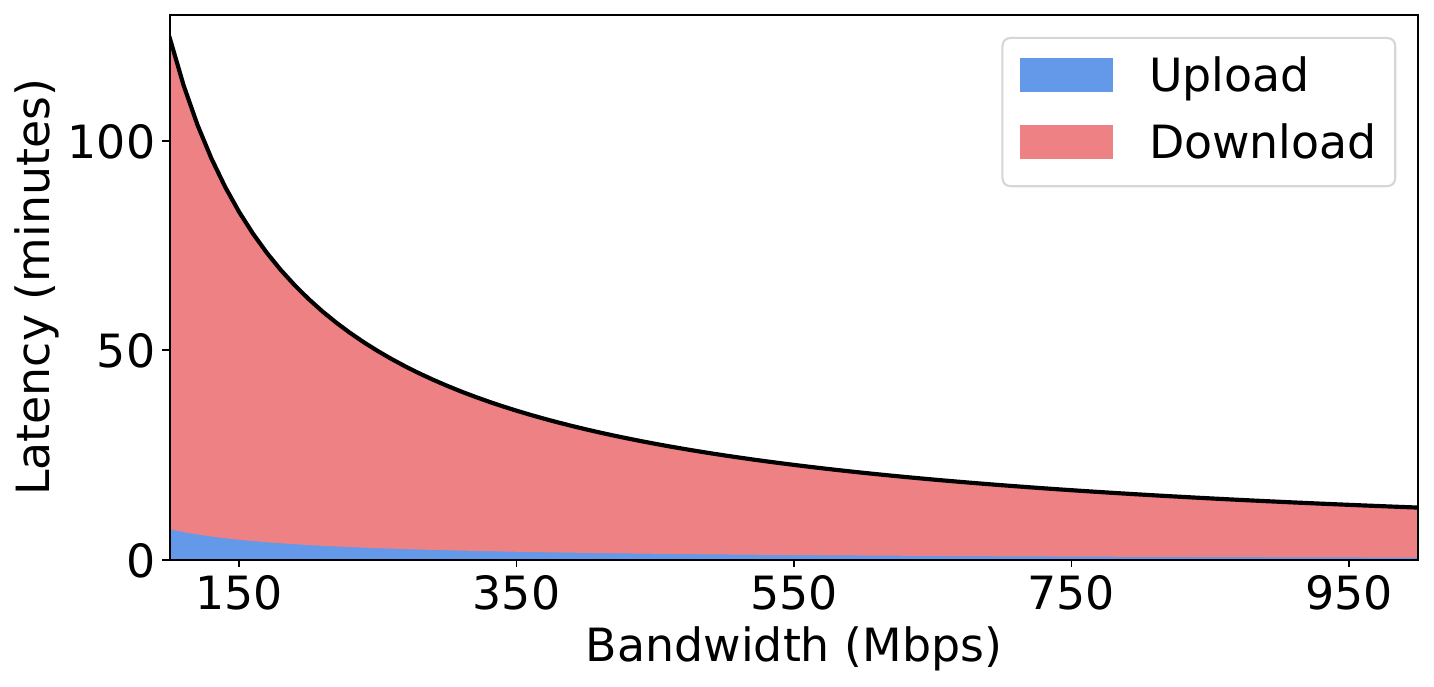} }}
    \vspace{-2em}
    \caption{Communication latency \textit{per inference} on ResNet-18 with TinyImageNet inputs as a function of total bandwidth. GC communication accounts for a majority of this latency.}%
    \label{fig:communication_issue}%
\end{figure}

Figure~\ref{fig:communication_issue} shows the total communication latency of PI for ResNet-18 on TinyImageNet. 
Bandwidth sensitivity is shown by sweeping total wireless capacity from \SI{100}{Mbps} to \SI{1}{Gbps}; the results assume an even split between the upload and download of the total bandwidth.
The data highlights a stark imbalance between the amount of information sent to the server (upload) and the amount of information received from the server (download). 
Most of the data transmission (81.5\% of the total) is in download due to the server sending GCs to the client device during its offline phase. (Recall that the ReLU GCs are on the order of tens of GBs for TinyImageNet).
Assuming a state-of-the-art 5G connection of \SI{1}{Gbps}, we can see the total communication time is 11 minutes, which is on par with the total HE computation costs.
The data also shows that much of the available wireless bandwidth is wasted: 50\% of the \SI{1}{Gbps} is reserved for upload even though upload accounts for only $5.7\%$ percent of the total transmitted data.
In Section~\ref{sec:optimizations} we show how better provisioning the wireless bandwidth can provide significant improvement.


\begin{table}[t]
\centering
\caption{Total time (seconds) for the  Server-Garbler protocol running ResNet-18 on TinyImageNet.}
\begin{tabular}{|c|c|c|c|c|c|}
\hline
        & GC   & HE   & SS    & Comms & Total \\ \hline
Offline & 25.1 & 1080 & 0.00  & 704   & 1809  \\ \hline
Online  & 200  & 0.00 & 0.610 & 42.5  & 243   \\ \hline
Total   & 225  & 1080 & 0.610 & 747   & 2052  \\ \hline
\end{tabular}
\label{tab:sg_total_time}
\end{table}

\begin{figure*}[!ht]%
    \centering
    \includegraphics[width=0.99\textwidth]{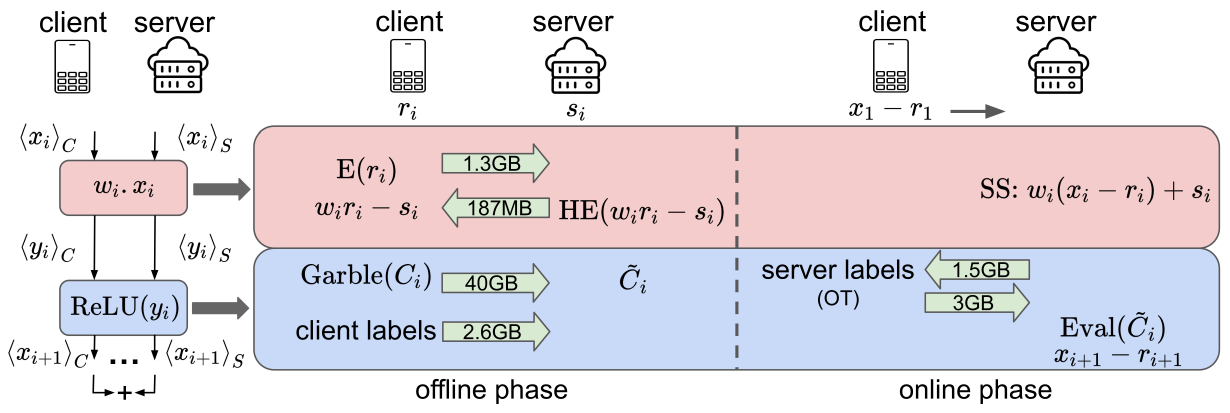}
    \caption{Proposed Client-Garbler protocol. The server (rather than the client) stores the garbled circuits ($\tilde{C})$. In addition, server labels must now be communicated via OT in the online phase since they are input dependent.}
    \label{fig:client-garbler}
\end{figure*}

\begin{figure}[!t]%
    \centering
    {{\includegraphics[width=0.99\columnwidth]{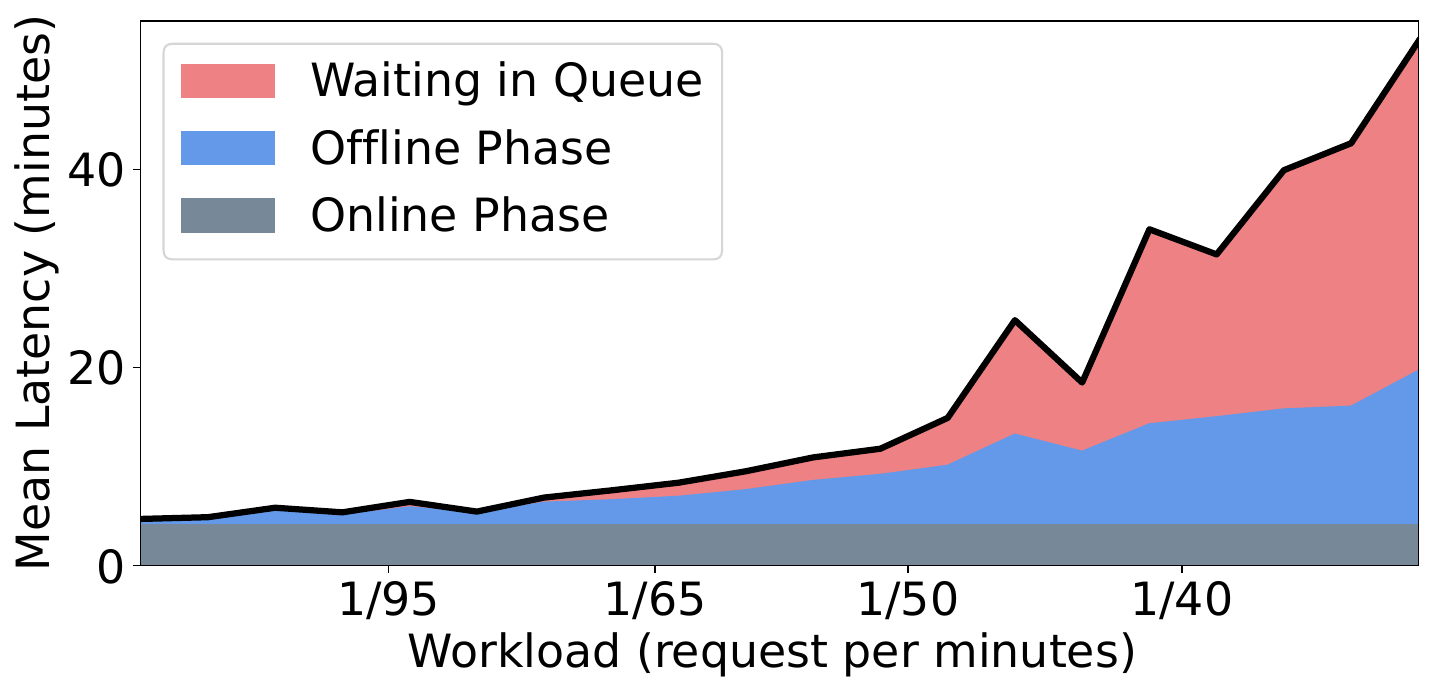} }}
    \vspace{-2em}
    \caption{Mean PI Inference latency with increasing inference requests rate (requests/min).}%
    \label{fig:non_zero_arrival_rates}%
\end{figure}

\subsection{PI with Streaming Inference Requests}

Table~\ref{tab:sg_total_time} presents the total PI time breakdown for a single ResNet-18 inference on TinyImageNet.
Prior work has assumed that the entirety of the offline latency, accounting for 88\% of the total inference run-time, can be hidden and assumed to have completed sometime prior to when the actual inference must take place.
While this may be possible when computing a single inference, it is not the case when considering a rate of inferences to process.

The fundamental problem that has been largely overlooked is that the storage costs are so extreme that systems can realistically only store pre-computes for a few inferences, if any.
These pre-computes serve as the buffer between online and offline cost.
As requests start to arrive, this buffer is rapidly depleted, as each takes GBs of storage limiting what the client can hold.
And even though the offline phase can run in parallel with the online phase as requests arrive,
the high offline latency (1808 seconds) can only be hidden at extremely low arrival rates.
In this light, the performance claims of hybrid systems, which consider isolated inferences and optimize only for online latency, have been overly optimistic and sweep large overheads (up to $88\%$ of end-to-end latency) under the rug.

In Figure~\ref{fig:non_zero_arrival_rates} we vary the arrival rate of inference requests from 1 request every 3 hours to 1 request every 15 minutes for ResNet-18 on TinyImageNet. 
We assume a generous \SI{128}{GB} of client-side storage is available to hold pre-computes.
The workload is simulated for 24 hours and we average the inference latencies over 50 runs of the simulation where each run is a unique sample from the arrival rate distribution. 
At the near-zero arrival rate region (far left of Figure~\ref{fig:non_zero_arrival_rates}), the mean inference latency is attributed completely to the online phase (grey) of the PI Protocol, as expected.
As the arrival rate increases, and starting at 1 request every two hours, mean average latency is impacted by the offline phase (blue), implying that the arrival rate has surpassed the rate of pre-compute refill.
And at 1 request every 70 minutes, we observe inference rates beginning to queue up (red). By one request every 30 minutes,  each inference must pay the full PI latency (online+offline) per request in addition to waiting in the queue for previous requests to complete. As request pile up, we see the queue waiting time increase and quickly dominate the total latency.





\section{Proposed PI Optimizations}
\label{sec:optimizations}

This section presents our proposed optimizations for PI: 
the Client-Garbler protocol, layer-parallel HE, and wireless slot allocation.
We explain and quantify the benefit of each optimization and conclude with an evaluation against prior work to demonstrate the total improvement.

\subsection{The Client-Garbler Protocol}
\label{subsec:cg}
As previously shown, the client-side storage requirements of existing protocols are prohibitively high, reaching \SI{41}{GB} for a single ResNet-18 inference on TinyImageNet. 


\textbf{Solution:}
To overcome this limitation, we propose the Client-Garbler protocol.
Here, the GC roles of the client and server are reversed: the client becomes the garbler and the server the evaluator (see Figure~\ref{fig:client-garbler}). 
In the offline phase, the client garbles ReLUs and sends the GCs along with the client labels to the server which stores them for use during the online phase. 
The parties engage in base OT offline so that in the online phase, the server can obtain their inputs to the GC using extended OT. 
The function to be garbled remains exactly the same as before: combine the shares of the prior linear layer, perform ReLU, and mask the output with the client's random share for the following linear layer (as shown in Figure 5 of \cite{juvekar2018gazelle}). 
Furthermore, there are no security implications of this optimization since the security guarantees of GC hold regardless of the role each party plays.

\begin{figure}[!t]%
    \centering
    {{\includegraphics[width=0.99\columnwidth]{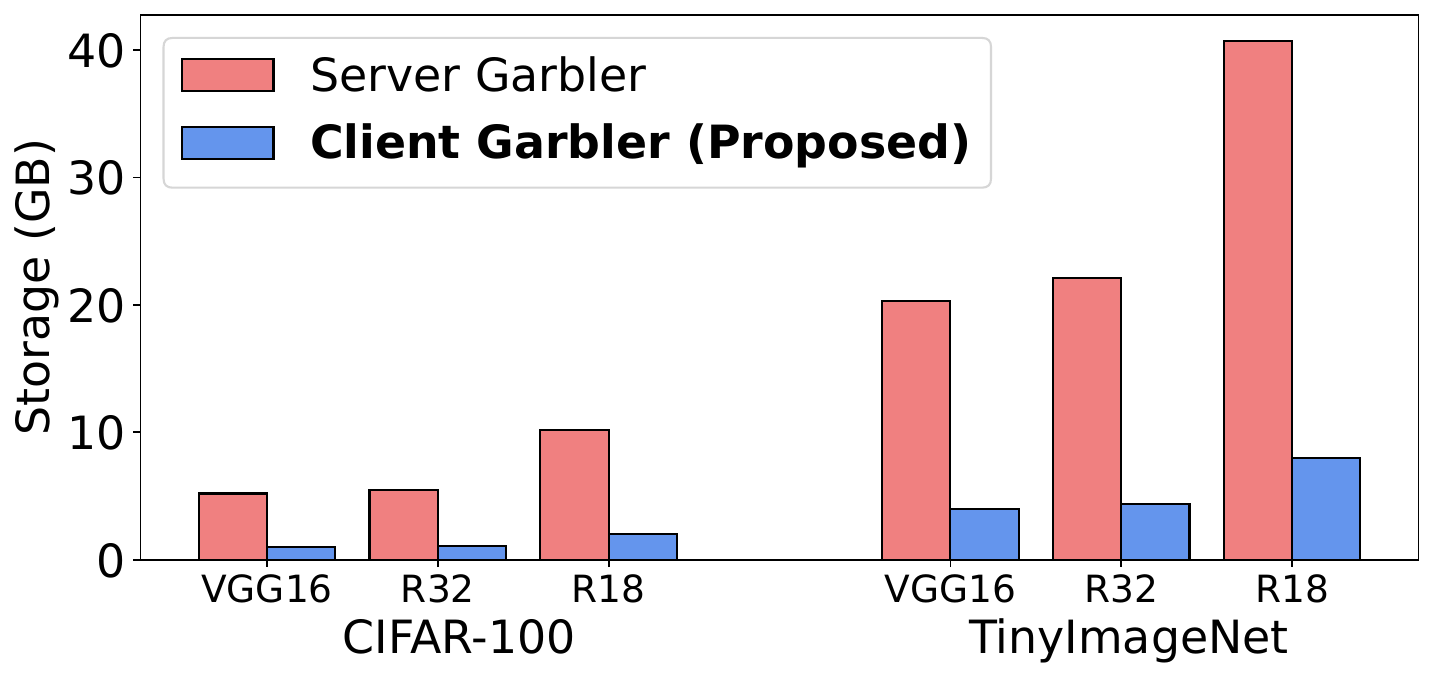} }}
    \vspace{-2em}
    \caption{Client-side storage requirements for baseline Server-Garbler and proposed Client-Garbler protocols.}%
    \label{fig:optimization_storage}%
\end{figure}
Using the Client-Garbler protocol, the large client-side storage requirements are now shifted to the server, thus taking advantage of the server's ample storage resources. Furthermore, the Client-Garbler protocol also reduces the online phase latency as the powerful server, rather than the client, performs GC evaluations.
A drawback is that the parties must now engage in OT during the online phase;
however, we observe OT communication time to be minimal and that Client-Garbler provides a net benefit to reducing online latency. 
In particular, the server-side evaluation of garbled ReLUs in the online phase outweighs the added communication latency from OT.
For example, on TinyImageNet on ResNet-18, Server-Garbler (Client-Garbler) has an online communication latency of 27.1 seconds (101 seconds) and a GC evaluation time of 200 seconds (11.1 seconds).
Thus, Client-Garbler provides an online speedup of 2.02$\times$.
The Client-Garbler also increases the offline run-time since GC garbling now runs on the slower client device. 
However, this is acceptable because 
the ability to store more pre-computes and mask offline costs altogether outweighs the increase in offline compute latency.

We evaluate the energy implications of switching GC roles using \textbf{\texttt{powertop}} \cite{powertop} to measure the energy consumption of garbling and evaluating 10,000 ReLUs (averaged over 30 independent runs on the Atom board).
Evaluation and garbling consume 1.25 and 2.33 Joules, respectively.
Thus, the Client-Garbler protocol increases client energy consumption by 1.8$\times$ as garbling requires additional encryptions compared to evaluating (Figure 1 in \cite{guo_garbling}). 
The compute latencies of both roles are also non-negligible (see Figure \ref{fig:compute_issue}).
These energy and performance profiles motivate research into custom hardware to accelerate both roles of GCs (discussed further in Section \ref{sec:discussion}.2). 
As noted in FASE~\cite{hussain2019fase}, accelerators can be built for both garbling and evaluation, as the computations are similar, and one simply has additional encryptions.

Nonetheless, the storage constraints of the Server-Garbler protocol prohibit the parties from engaging in an offline phase, leading to a significant increase in PI latency. 
For example on ResNet-18 on TinyImageNet, with \SI{16}{GB} client storage, the Client-Garbler has an online latency of 2.3 minutes while the Server-Garbler's is 34.0 minutes. 
Thus, the Client-Garbler protocol results in a significant decrease in PI latency although the client consumes more GC energy.


In the absence of a full-system perspective, i.e., ignoring storage constraints, the asymmetry between the client and server, and assuming single inferences, Server-Garbler is indeed a natural choice since it moves OT to the offline phase and reduces client-side energy consumption.
Accounting for these factors, however, we find that Client-Garbler significantly outperforms Server-Garbler.


\textbf{Benefits of Client-Garbler:}
Figure~\ref{fig:optimization_storage} highlights the benefits of the Client-Garbler protocol over the Server-Garbler protocol in terms of client-side storage. On average we find the optimizations reduces the storage requirement of the client by $5\times$. As a concrete example, the Client-Garbler protocol reduce the client-side storage footprint of ResNet-18 on TinyImageNet from \SI{41}{GB} to \SI{8}{GB}. 
A discussion of the benefits of Client-Garbler on mean PI latency is deferred to Section~\ref{sec:alltogether}.



\subsection{Layer-Parallel Homomorphic Encryption (LPHE)}
In order to sustain high inference request arrival rates using PI, it is crucial to not only optimize the online phase but also the offline latency. 
During the offline phase, hybrid protocols incur a large HE run-time in order to generate additive secret shares for use during the online phase. For example, the HE run-times for ResNet-18 on TinyImageNet is $17.76$ minutes, which makes up over $80\%$ of the entire protocol's compute latency.

\textbf{Solution:} We identify a novel parallelization opportunity for HE, when used in hybrid, protocols termed layer-parallel HE (LPHE). 
In hybrid protocols, HE is used server-side to securely evaluate the linear layers of the neural network on the client's randomly sampled secret shares (as shown in Figure \ref{fig:pi_protocol}). We observe that these secret shares are \textit{generated independently} for each linear layer of the network. 
This means that the secure evaluation of each linear layer need not be performed in-order and can be run in an embarrassingly parallel manner on the server.
LPHE provides a general mechanism for speeding up HE computations as a function of the number of layers in the network. 

\begin{figure}[!t]%

    \centering
    {{\includegraphics[width=0.99\columnwidth]{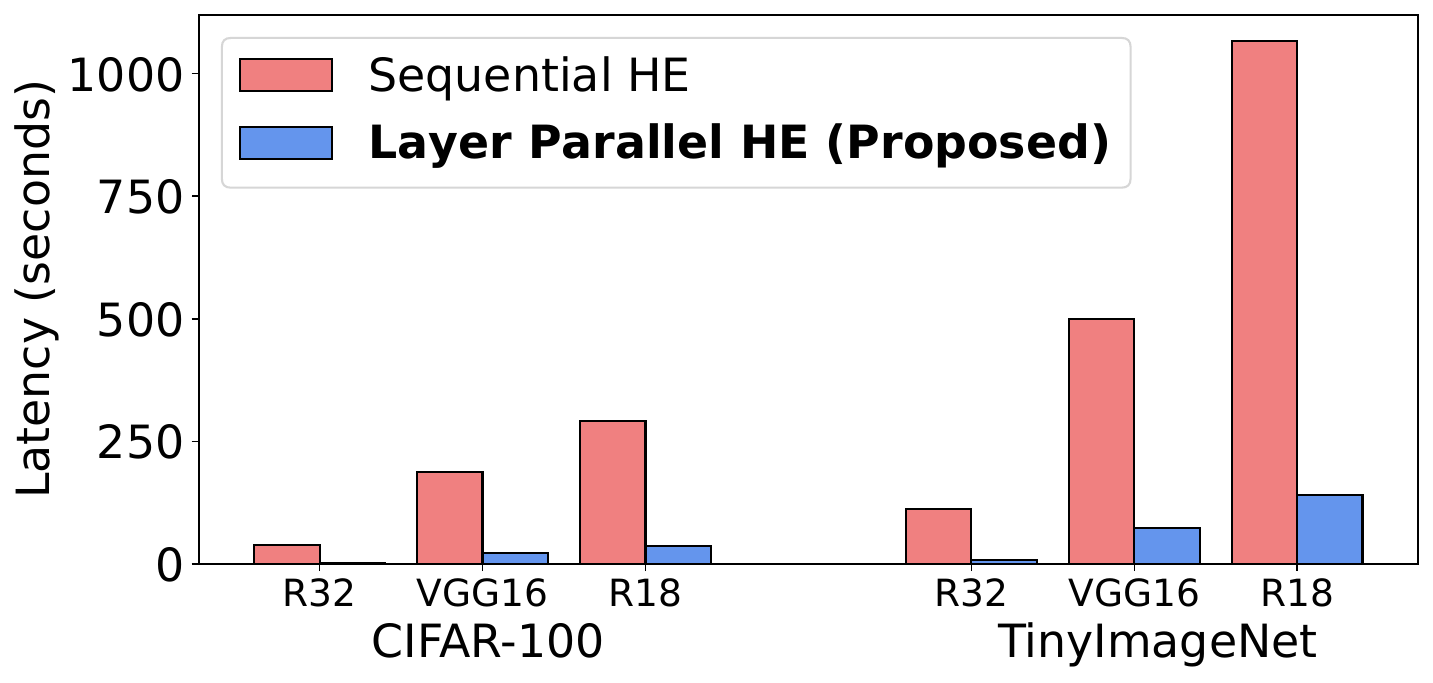} }}
    \vspace{-1em}
    \caption{Baseline sequential vs. proposed layer parallel HE (LPHE) latency on the server.}%
    \label{fig:optimized-compute}%

\end{figure}


\begin{figure*}[!t]%
    \centering
    \subfloat[Client-Side Storage: 8GB, 16GB]{{\includegraphics[width=0.66\columnwidth]{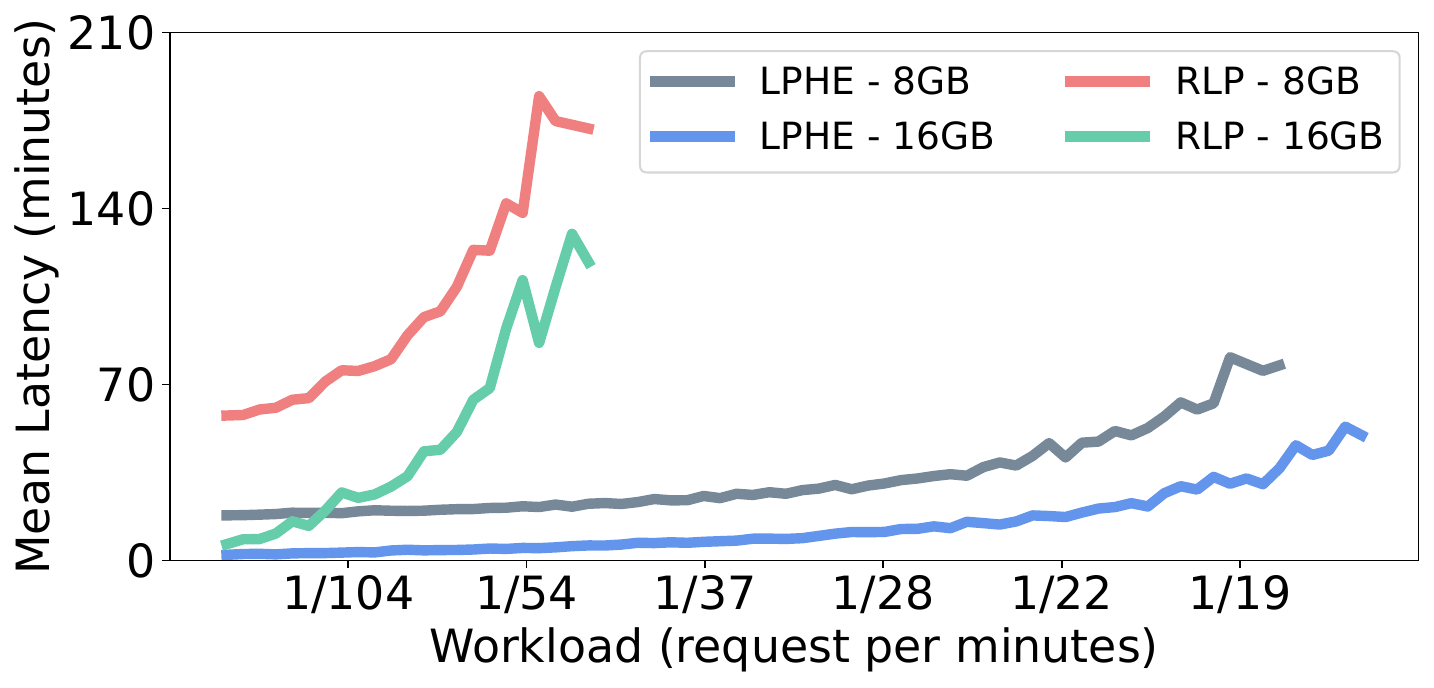} }}
     \subfloat[Client-Side Storage: 32GB, 64GB]{{\includegraphics[width=0.66\columnwidth]{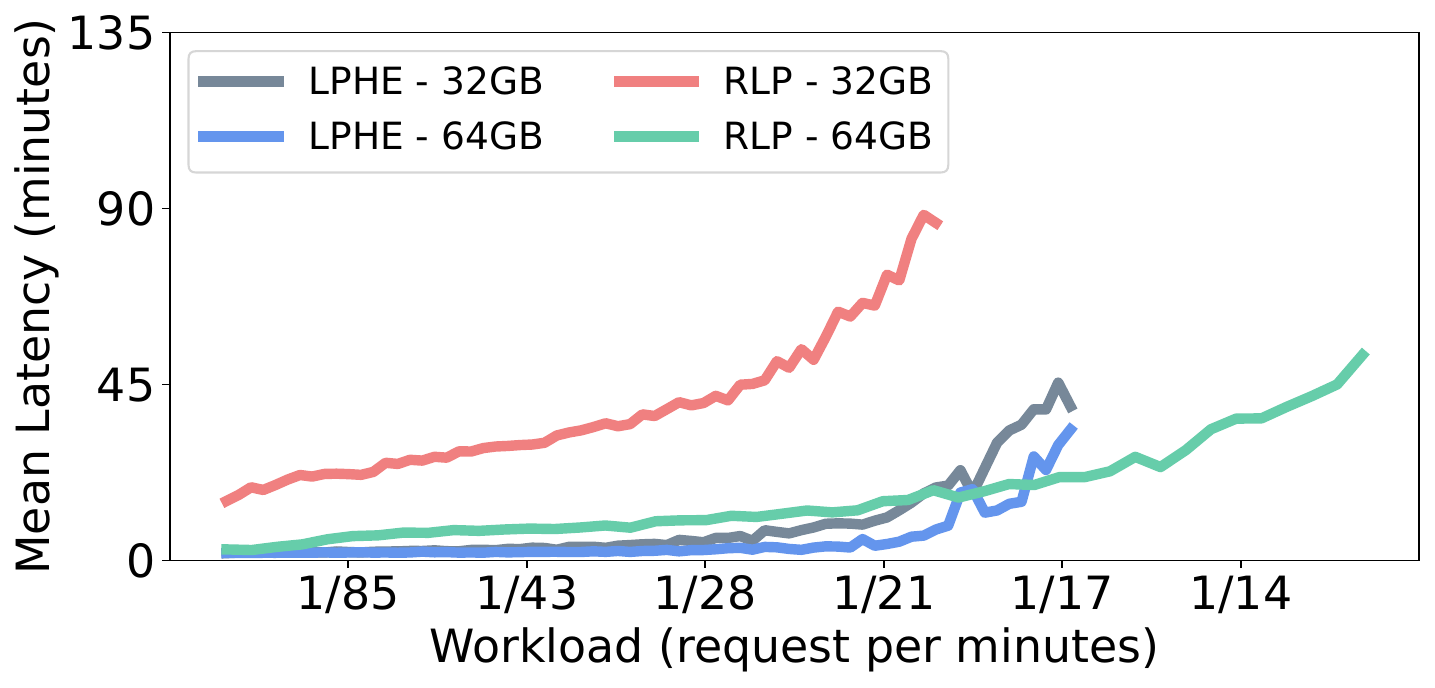} }}
    \subfloat[Client-Side Storage: 140GB]{{\includegraphics[width=0.66\columnwidth]{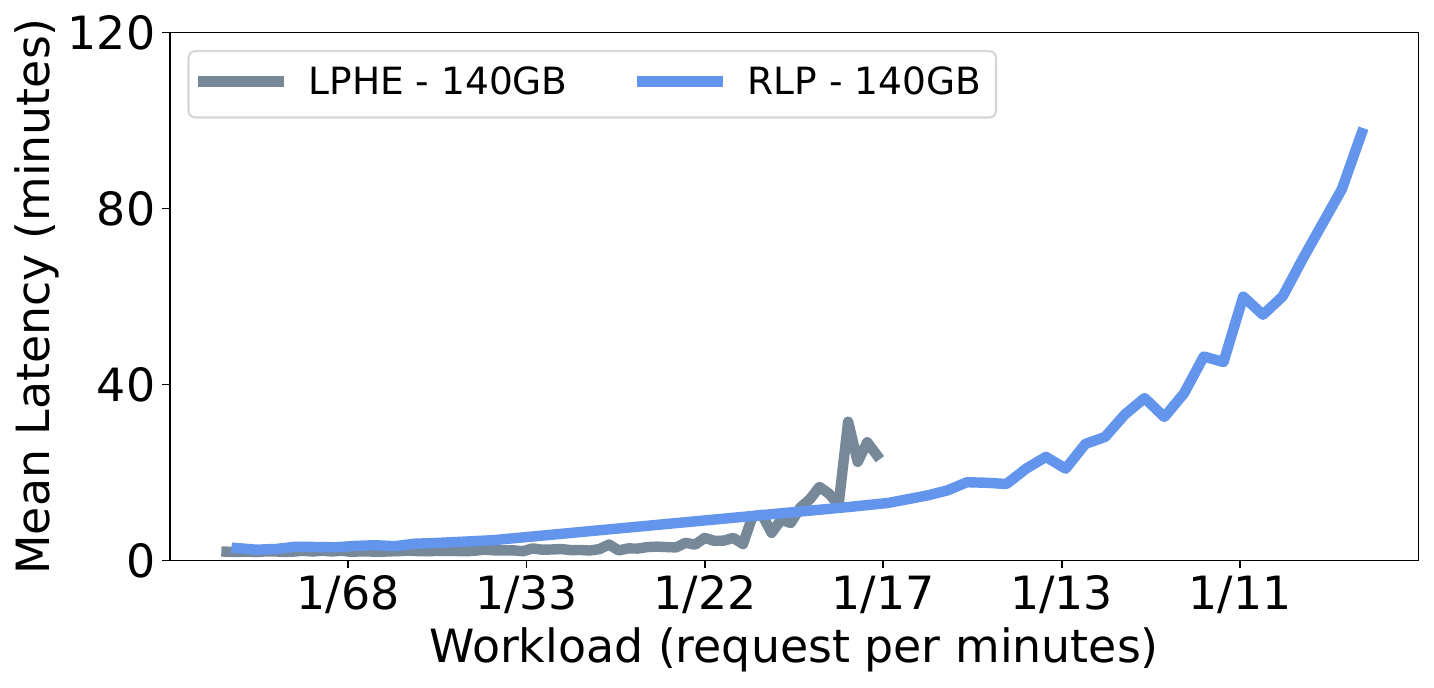} }}
    \vspace{-.5em}
     \caption{LPHE versus Request-Level Parallelism (RLP) for low (a), medium (b), and high (c) client-side storage. At 140GB, both LPHE and RLP utilize the same amount of server and client resources.} 
    \label{fig:lphe_vs_rlp}%

\end{figure*}

In PI protocols like Gazelle, which do not use a combination of SS and HE for linear layers, HE evaluation is performed directly on network inputs to linear layers and LPHE cannot be leveraged as layers must be processed sequentially. 
Thus, PI protocols that combine HE and SS \emph{can be faster than protocols that use HE alone.}
This is counter-intuitive as HE$+$SS protocols are more complex and require more work.
However, the opportunity to parallelize HE across layers more than makes up for the extra SS work. 

\textbf{Benefits of LPHE:}
Figure~\ref{fig:optimized-compute} shows the performance benefit of LPHE compared to prior work. 
The impact of LPHE is substantial on the overall HE speedup and we observe that the LPHE run-time is bounded by the longest-running HE linear layer.
Therefore, we expect LPHE to scale well for even deeper networks. For ResNet-18 on TinyImageNet, LPHE reduces the HE evaluation run-time from $17.76$ minutes to just $2.35$ minutes. Across all datasets and networks, LPHE speeds up HE by $9.7\times$.

Several prior works have optimized HE leveraging the available parallelism within the core kernels and across ciphertexts \emph{within} a layer, often for significant speedup as custom hardware accelerators~\cite{reagen2021cheetah,samardzic2021f1,samardzic2022craterlake}. 
We note that LPHE is a completely orthogonal source of parallelism, and it can be combined with the prior work for even more performance.
These recent papers report that HE computations can be brought within a constant factor of plaintext performance; 
LPHE could be a significant step towards closing the remaining performance gap.

\textbf{Comparison with Request-Level Parallelism:}
\label{subsub:lphe}
LPHE distributes the independent HE layer computations for a single inference pre-processing task across available cores.
An alternative approach is for each core to handle independent inference pre-processing tasks, which we call Request-Level Parallelism (RLP). 
We now explore the trade-offs of LPHE and RLP in the context of varying the client-side storage capacity and inference workload for our proposed protocol (Client-Garbling and wireless slot allocation discussed in Section \ref{subsec:wsa}) on the TinyImageNet dataset for ResNet-18.
For RLP, a single core is used on both devices to run a pre-processing phase.
Thus, if the client-side storage is sufficient to buffer $k$ inference precomputes, the two parties can engage in $k$ pre-processing phases concurrently. 

Figure \ref{fig:lphe_vs_rlp} presents the results of our experiments and assumes 17 cores for server-side HE (there are 17 linear layers in ResNet18) and 4 cores for the client-side GC evaluation.
At \SI{8}{GB} of client storage (Figure \ref{fig:lphe_vs_rlp}a), pre-processing cannot be engaged for either LPHE or RLP and the entire PI protocol must be processed online.
Here, LPHE leverages all 17 server cores for an end-to-end latency of 1053 seconds. 
Meanwhile RLP is limited to a single core, running in 3126 seconds. 
With \SI{16}{GB} (Figure \ref{fig:lphe_vs_rlp}a), both RLP and LPHE inference latency improve as the pre-processing phase can complete offline during the downtime between requests. 
However, RLP can only engage in a single pre-processing phase, given storage constraints, and under-utilizes the cores, taking 3013 seconds per inference compared to LPHE's 936 seconds. 
We repeat these experiments using 32 and \SI{64}{GB} (Figure \ref{fig:lphe_vs_rlp}b) for completeness, as these were assumed in other experiments. 
Here, the client device can store 3 and 7 precomputes, respectively, and RLP is better able to leverage the server resources with independent pre-processing.
In the case of \SI{32}{GB}, RLP now sustains a higher workload at 1 request per 21 minutes, but still has a higher mean inference latency when compared to LPHE. 
When the client device has \SI{64}{GB} for precomputes, RLP is able to handle a higher workload (1 request per 13 minutes) compared to LPHE (1 request per 17 minutes), since RLP has a higher throughput for precomputes than LPHE.
Finally, we set the client-side storage to \SI{140}{GB} (Figure \ref{fig:lphe_vs_rlp}c); this allows for the client to store 17 precomputes and so RLP and LPHE utilize the same number of resources on both devices: all 17 server cores and all 4 client cores.
Now, RLP can handle a workload of 1 request per 10 minutes while LPHE sustains a maximum workload of 1 request per 17 minutes, since RLP has a higher pre-compute throughput (1 precompute every 627 seconds) compared to LPHE (1 precompute every 939 seconds).

Without restricting storage, RLP results in a higher throughput of precomputes compared to LPHE since RLP distributes an equal amount of work across available cores. Meanwhile, LPHE distributes an unbalanced amount of work across cores since each linear layer’s homomorphic evaluation varies in latency, thus leaving cores partially under-utilized. 
However, RLP only realizes its potential when the client is able to devote an immense amount of storage to precomputes, which is unlikely given the dozens to hundreds of GBs required and limited storage of clients' devices, e.g., a smartphone. 
Moving forward it is likely that the two approaches will be combined and adapt to the available storage to maximize throughput while minimizing latency.
\begin{figure}[!t]%
    \centering
    {{\includegraphics[width=0.99\columnwidth]{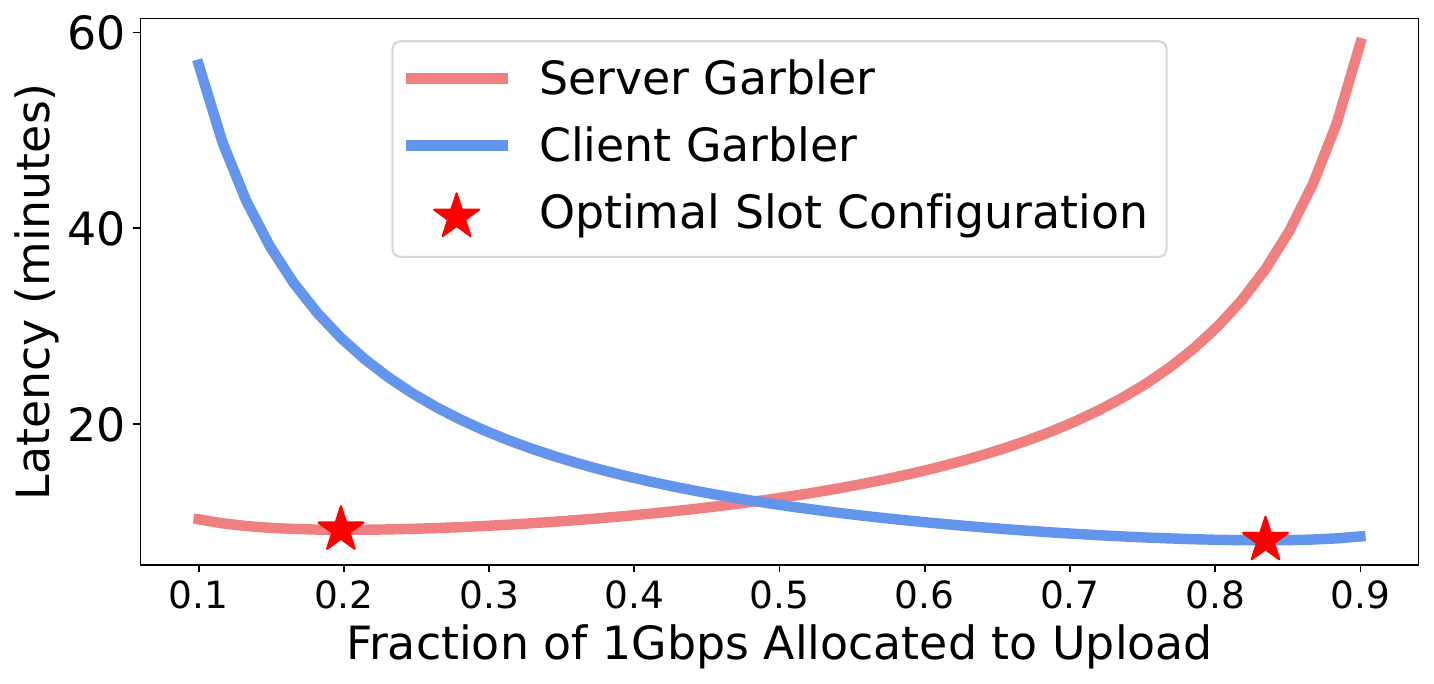} }}
    \vspace{-2em}
    \caption{Impact of wireless slot allocation (WSA) on communication latency for Server-Garbler and Client-Garbler protocols. Optimal WSA points that minimize communication for the two protocols are highlighted.}
    \label{fig:optimization_comz}
\end{figure}
\begin{figure*}[!t]%
    \centering
    \subfloat[CIFAR-100, ResNet-32]{{\includegraphics[width=0.66\columnwidth]{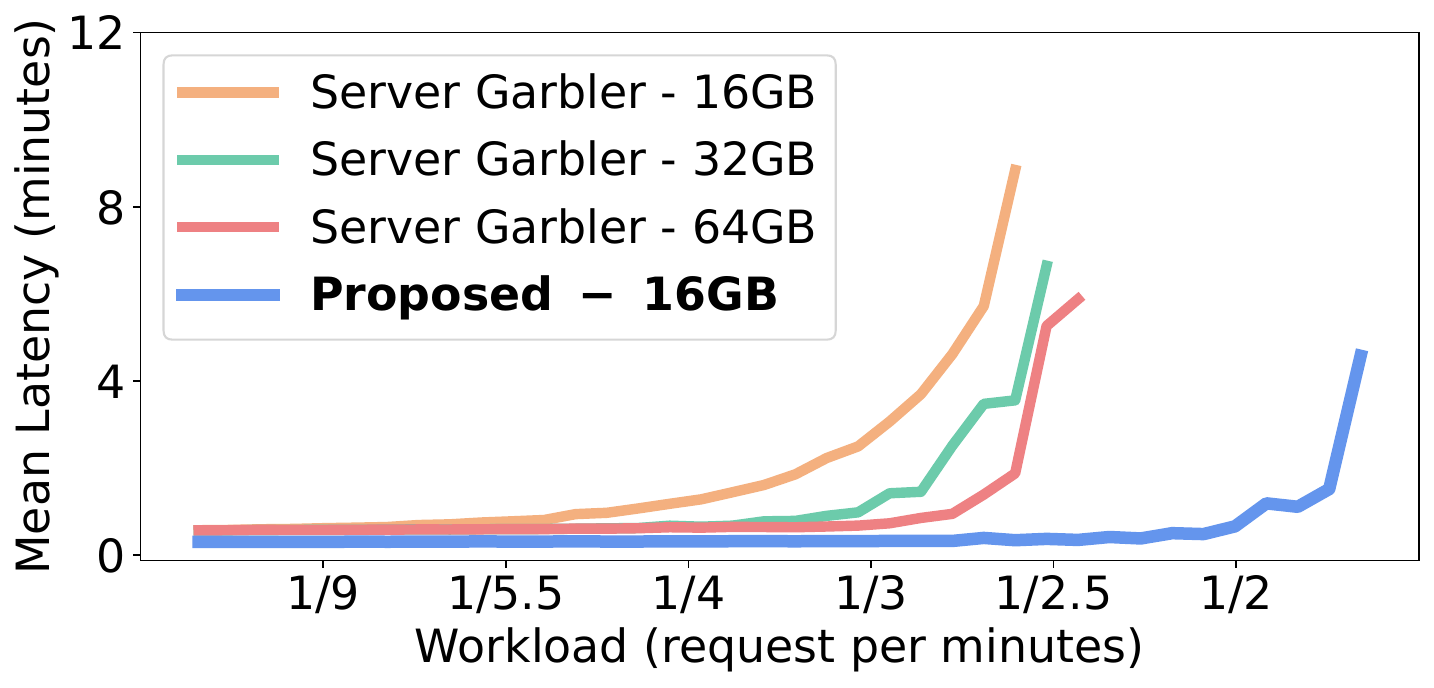} }}
     \subfloat[CIFAR-100, VGG-16]{{\includegraphics[width=0.66\columnwidth]{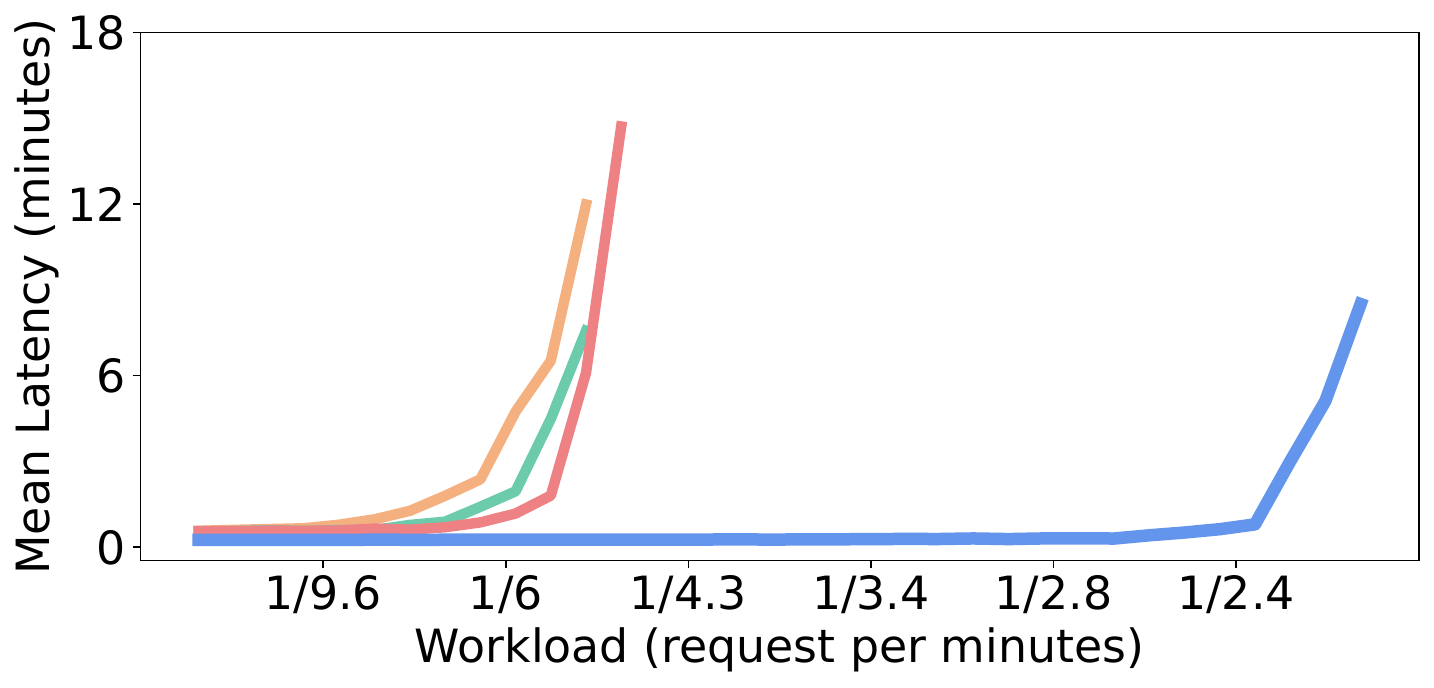} }}
    \subfloat[CIFAR-100, ResNet-18]{{\includegraphics[width=0.66\columnwidth]{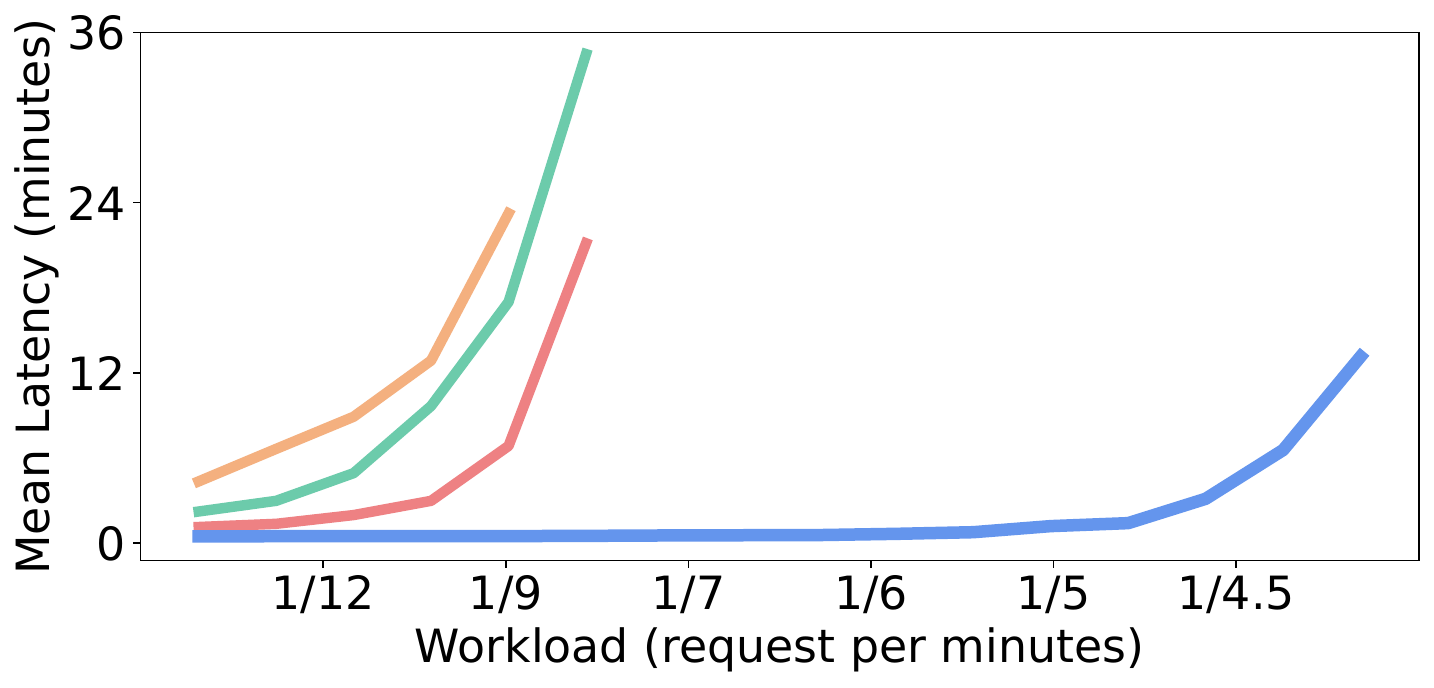} }}
    \\
    \subfloat[TinyImageNet, ResNet-32]{{\includegraphics[width=0.66\columnwidth]{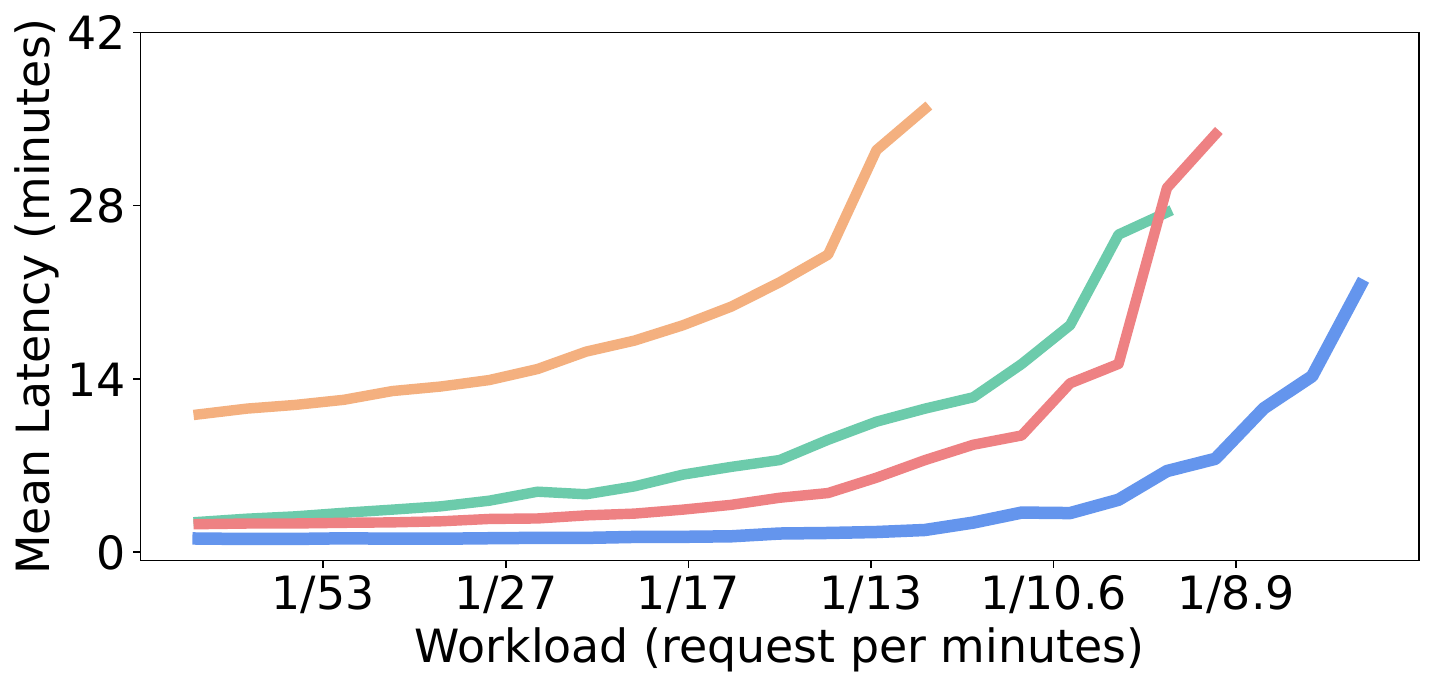} }}
    \subfloat[TinyImageNet, VGG-16]{{\includegraphics[width=0.66\columnwidth]{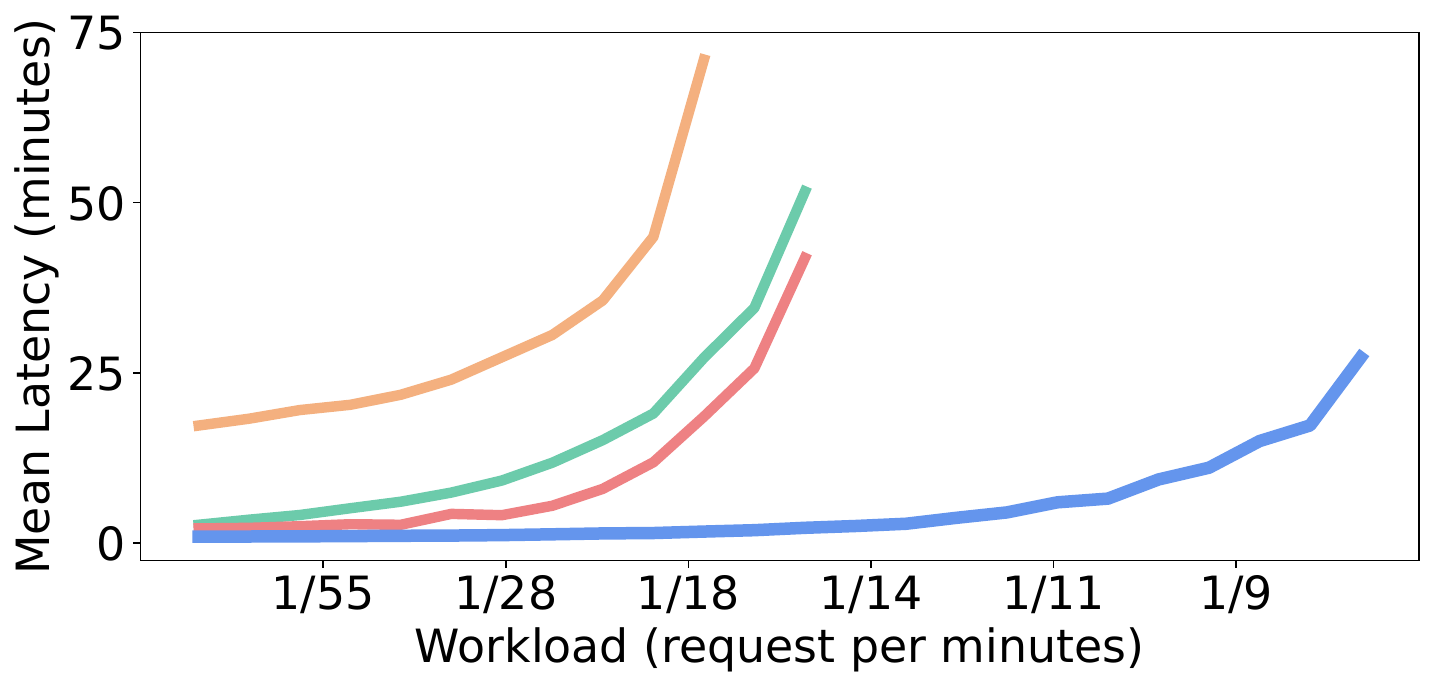} }}
    \subfloat[TinyImageNet, ResNet-18]{{\includegraphics[width=0.66\columnwidth]{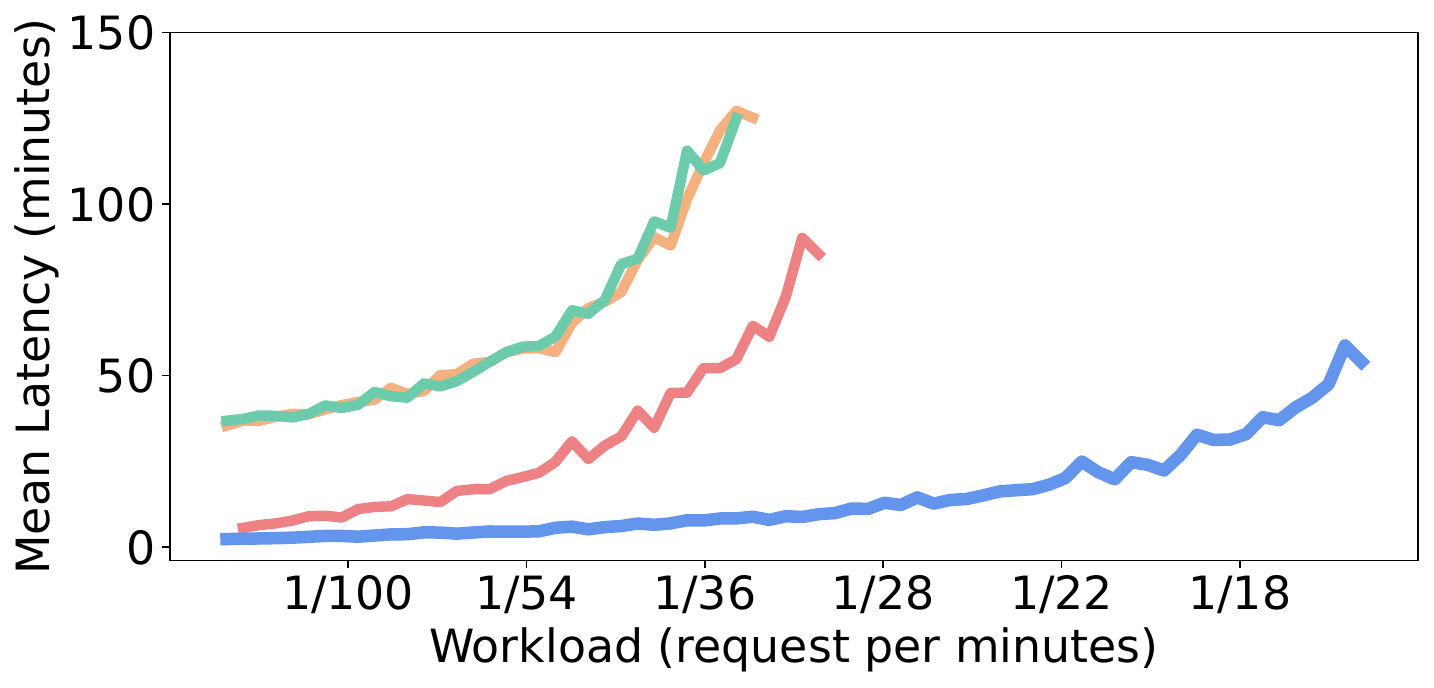} }}

    \caption{Comparison of baseline Server-Garbler and proposed optimizations. In each case, the proposed optimizations (with limited client-side storage) exhibit a lower mean inference latency and sustain a higher arrival rate.} 
    \label{fig:twostacked}%

\end{figure*}
RLP also provides benefits in a setting where multiple clients use a single server to process their independent requests. Here, the total client storage scales with the number of clients (i.e., the number of requests). For example, if there were 9 clients each with \SI{16}{GB} (as in Figure \ref{fig:lphe_vs_rlp}a), there would be a total of \SI{144}{GB}  net client storage, similar to the assumption in Figure \ref{fig:lphe_vs_rlp}c (\SI{140}{GB}). As a result, the server can exploit RLP and sustain a high throughput without needing LPHE. 
However, each client still has only enough storage for a single precompute, (which must be retained on the client device)---hence the cost of performing this precompute will still be incurred online at higher arrival rates, increasing client latency. From the perspective of each client, the latency would still be similar to the \SI{16}{GB} results shown in Figure \ref{fig:lphe_vs_rlp}a.


\subsection{Wireless Slot Allocation}
\label{subsec:wsa}
Our final optimization seeks to reduce the communication latency of hybrid PI protocols.
As we observed in Section~\ref{sec:characterization}, there is significant asymmetry in the amount of data sent from the client to server and vice versa.
Therefore, we propose to minimize communication latency by matching the upload and download bandwidths to the data transferred from client to server and server to the client, respectively. 
Fortunately, 5G wireless standards provide exactly this flexibility.


\textbf{Solution:} Current 5G wireless standards use time division duplexing (TDD) to partition bandwidth between upload and download~\cite{3gpp20153gpp}. 
A \SI{10}{ms} frame is divided into 10 sub-frames, each of which can be allocated to either upload or download, allowing significant flexibility in the fraction of bandwidth allocated in each direction. This flexibility can be, and has been, used to support autonomous driving and virtual reality applications, for instance, where almost all communication is in the upload~\cite{5G_TDD_Uplink_WhitePaper}.
The sub-frame structure can also be dynamically changed at a ms granularity~\cite{tang2020deep}.
We propose to exploit the flexibility in 5G wireless standards to match the asymmetric communication between client and server in hybrid PI protocols, allocating more upload slots for the Client-Garbler protocol and more download slots for Server-Garbler.

\textbf{Benefit:}
Figure~\ref{fig:optimization_comz} shows the benefit and potential of the wireless slot allocation (WSA) optimization. Using our simulator, we set the total available bandwidth to \SI{1}{Gbps} and calculate the total offline and online communication latency for both protocols while sweeping the percentage of slots allocated to upload.
First, we note that as more slots are allocated for upload (download), we approach an optimal slot configuration for the Client-Garbler (Server-Garbler) protocol. 
Second, we find that an even bandwidth split works well as a starting point when compared to sub-optimal settings. 
However, selecting the optimal point can provide up to a $35\%$ communication time reduction. 

We note that the ideal bandwidth partitions are not perfectly symmetric across protocols: 
Server-Garbler is optimal with \SI{802}{Mbps}  configured for download bandwidth and Client-Garbler is optimal with \SI{835}{Mbps} set for upload bandwidth; this is due to the Client-Garbler protocol performing OT during the online phase.

\subsection{Putting it All Together}
\label{sec:alltogether}

\begin{figure*}[!t]%
    \centering
    \subfloat[AMD Server (\textbf{1x})]{{\includegraphics[width=0.66\columnwidth]{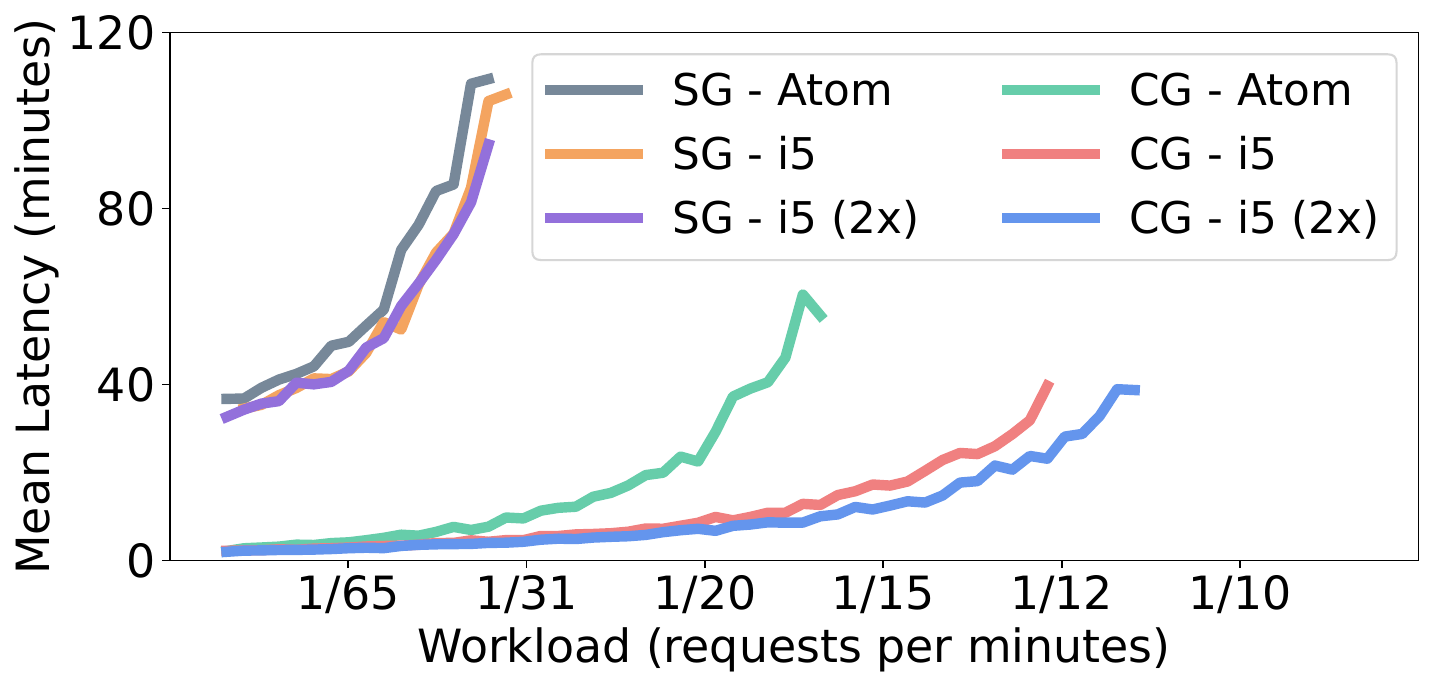} }}
     \subfloat[AMD Server (\textbf{2x})]{{\includegraphics[width=0.66\columnwidth]{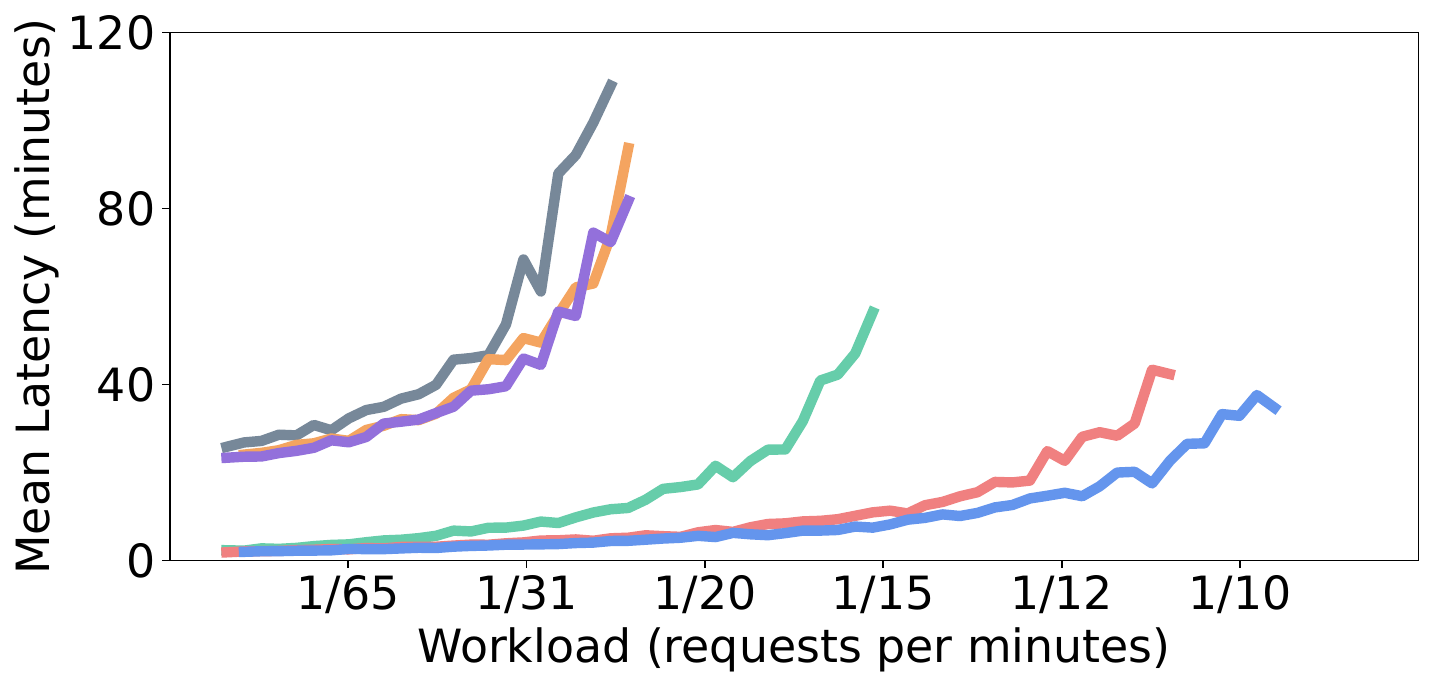} }}
    \subfloat[AMD Server (\textbf{4x})]{{\includegraphics[width=0.66\columnwidth]{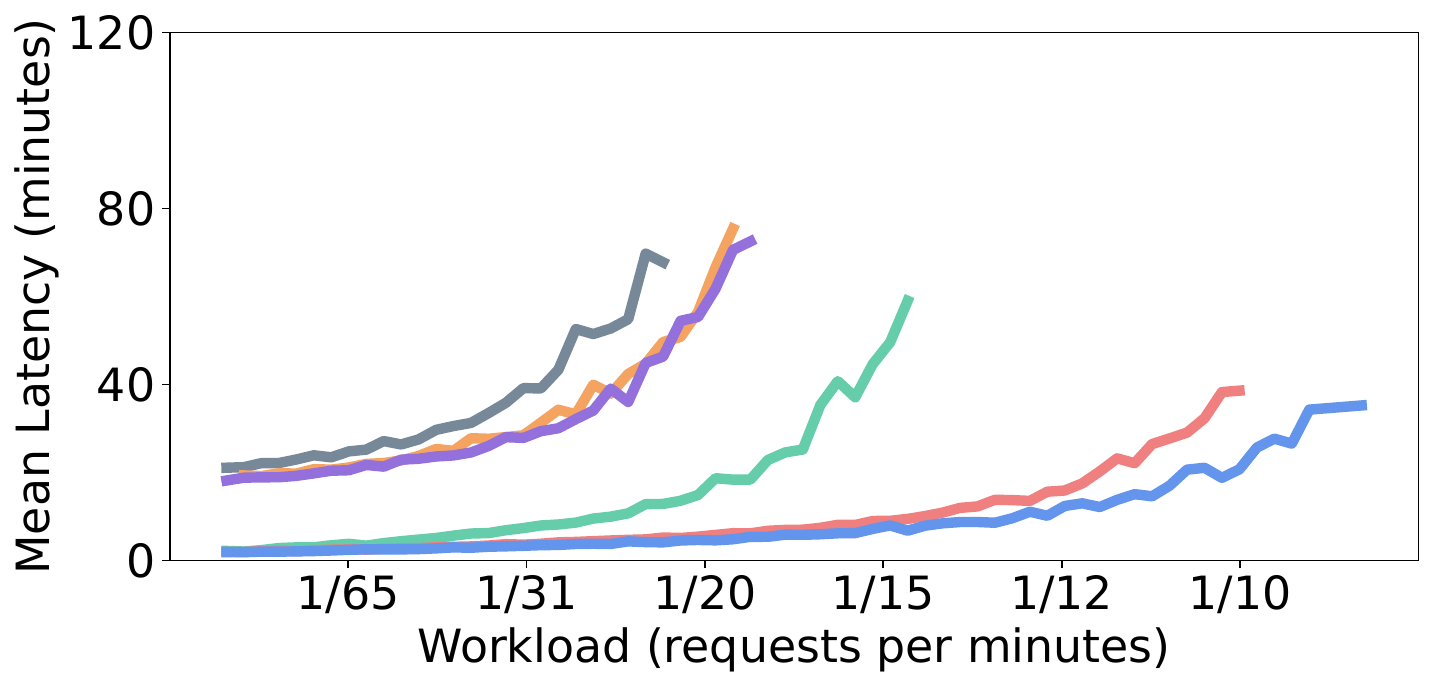} }}
    \vspace{-.5em}
     \caption{Sensitivity studies of server and client compute capabilities while fixing the client-side storage to 16GB for TinyImageNet on ResNet-18 for both Server-Garbler (labeled SG) and Client-Garbler (labeled CG).} 
    \label{fig:sensitivity_sweep}%

\end{figure*}

Here we evaluate a standard high-performance protocol, namely the Server-Garbler protocol, against our optimizations (Client-Garbler, LPHE-enabled, and WSA-optimal).
We analyze both protocols for all network-dataset pairs and set the client-side storage capacity to 16, 32, and \SI{64}{GB} for the Server-Garbler protocol. 
For our proposed Client-Garbler protocol we set the client-side storage capacity to the lowest setting (\SI{16}{GB}), as this is sufficient space to store at least a single inference pre-compute.
We set the server-side storage capacity to \SI{10}{TB} and the total available bandwidth to \SI{1}{Gbps}.
We sweep arrival rates to cover the online-only and maximal sustainable throughput regions of all workloads and simulate over a period of 24 hours.
For robust data, we calculate the mean inference over 50 independent runs of the simulation.
Figure~\ref{fig:twostacked} presents the results of these experiments.

For CIFAR-100 (Figure~\ref{fig:twostacked}a-c) both the Server-Garbler and our Client-Garbler are able to store at least a single inference pre-compute across all storage capacities considered and are thus able to engage in separate offline and online phases.
With Server-Garbler, the mean inference latency quickly rises to the maximum sustainable throughput (as seen by the asymptotes), even when increasing the client-side storage capacity.
This is caused by the inference request arrival rates outpacing the latency and resources required to engage in the offline phase of the Server-Garbler protocol and thus the costs must be incurred online. 

Compared to the Server-Garbler protocol, our optimizations maintain a lower mean inference latency due to reduced client-storage pressure and computation/communication latency.
Additionally, we observe that our proposed protocol has  lower latency than the Server-Garbler protocol in the low arrival rate region (e.g., 1.88 minutes on ResNet-18 for TinyImageNet) as the server performs GC evaluation rather than the client 
and there is sufficient time between requests to replenish pre-computes. 
Finally, we note that the maximum sustained throughput of the proposed protocol extends out towards heavier workloads and thus can sustain a higher arrival rate when compared to prior work. 
For ResNet-18 on CIFAR-100, our proposed protocol can handle a request rate of 1 every 5 minutes while the Server-Garbler protocol is capped to 1 inference request every 12 minutes.

When moving from CIFAR-100 to TinyImageNet (Figure~\ref{fig:twostacked}d-f), we first observe that not all client-side storage settings allow for the Server-Garbler protocol to engage in the pre-computation or the offline phase. 
Thus, the offline costs are entirely shifted online,
even for the low arrival rate region. For example on ResNet-18 with TinyImageNet, both the 16 and \SI{32}{GB} settings of the Server-Garbler are unable to engage in an offline phase as \SI{41}{GB} are required for storing GCs. Hence, both the 16 and \SI{32}{GB} configurations start with a high mean inference latency of nearly 1 hour. 
On the other hand, our proposed protocol can perform both the offline and online phase with just \SI{16}{GB} of storage allocated to the client. Finally, our Client-Garbler protocol is able to sustain higher arrival rates as both LPHE and optimal WSA reduce compute and communication latency.


\subsection{Sensitivity Study of Compute Capabilities}
\label{sec:sensitivity}

We now show the sensitivity of the Server-Garbler and proposed Client-Garbler protocol to client/server performance by sweeping device capabilities. 
We assume the following configurations for the client device: our baseline Intel Atom board, an Intel i5 processor, and a device with $2\times$ the compute capabilities of the i5. 
For the server: our baseline AMD EPYC 7502 32-core processor and servers with $2\times$ and $4\times$ the compute capabilities of the AMD chip (we initially benchmarked an Intel Xeon Gold 5218 CPU @ \SI{2.30}{GHz}, but the results were not significantly different from the AMD machine).
Figure \ref{fig:sensitivity_sweep} shows mean inference latencies for TinyImageNet on ResNet-18 with a client-side storage capacity of \SI{16}{GB}.

Figure \ref{fig:sensitivity_sweep}a uses the standard AMD processor for the server and varies the client device. 
For the Server-Garbler protocol, the lack of sufficient client-side storage to buffer precomputes increases the end-to-end latency across all sustainable workloads as the entire protocol is executed on-the-fly once an inference is requested. Irrespective of the client device, the Server-Garbler protocol sustains a maximum workload of roughly 1 request per 30 minutes. For the Client-Garbler protocol, the \SI{16}{GB} client-side storage is sufficient to buffer a single precompute; we observe a decrease in mean inference latency across all workloads as the offline phase can now be executed. 
When moving from an Atom to i5 client, the maximum sustainable throughput increases from 1 request every 15 minutes to 1 request every 10 minutes as client-side offline garbling reduces from 382.6 seconds (Atom) to 107.2 seconds (i5). 
A client device with $2\times$ the compute capabilities of the i5 chip helps increase the maximum sustainable workload by further decreasing the garbling latency to 53.8 seconds.

We observe similar trends when increasing the compute capabilities of our server to both $2\times$ and $4\times$ of our standard AMD server (Figure \ref{fig:sensitivity_sweep}b and Figure \ref{fig:sensitivity_sweep}c, respectively). 
The Server-Garbler protocol is still unable to engage in a precompute phase and must incur the entire PI protocol online. 
However, since server-side garbling and server-side HE evaluation latency is reduced, the Server-Garbler protocol sustains a higher workload extending to 1 request per 20 minutes in the case of AMD \textbf{$4\times$}. For the Client-Garbler protocol, a faster server reduces the server-side HE evaluations and server-side GC evaluation, thus reducing both the offline and online time: Client-Garbler extends its sustainable workload to 1 request per 9 minutes in the AMD \textbf{$4\times$} case.
Even in our optimal compute setting (i5 $2\times$ and AMD $4\times$), we observe minutes-long inference latencies, stressing the need for hardware accelerators to reduce the compute burden of both HE and GC.
An in-depth analysis of future optimization (including accelerators) for PI latency is shown in Section \ref{sec:discussion}.

\section{Discussion and Further Optimizations}
\label{sec:discussion}


In this section, we discuss opportunities for further improvement in PI latency that can 
be enabled by future research innovations.  To this end, we breakdown the total costs of 
Server-Garbler and Client-Garbler protocols to identify key bottlenecks, and show how these costs decrease with accumulating optimizations. The results are shown in Figure~\ref{fig:analysis} which plots total PI latency (top) and normalized latency broken down into various components (bottom).
The numbers on top of the normalized latency bars represent the fraction of PI latency incurred offline.

\subsection{Comparing Protocol Costs}
\begin{figure}[!t]%
    \centering
    {{\includegraphics[width=0.9\columnwidth]{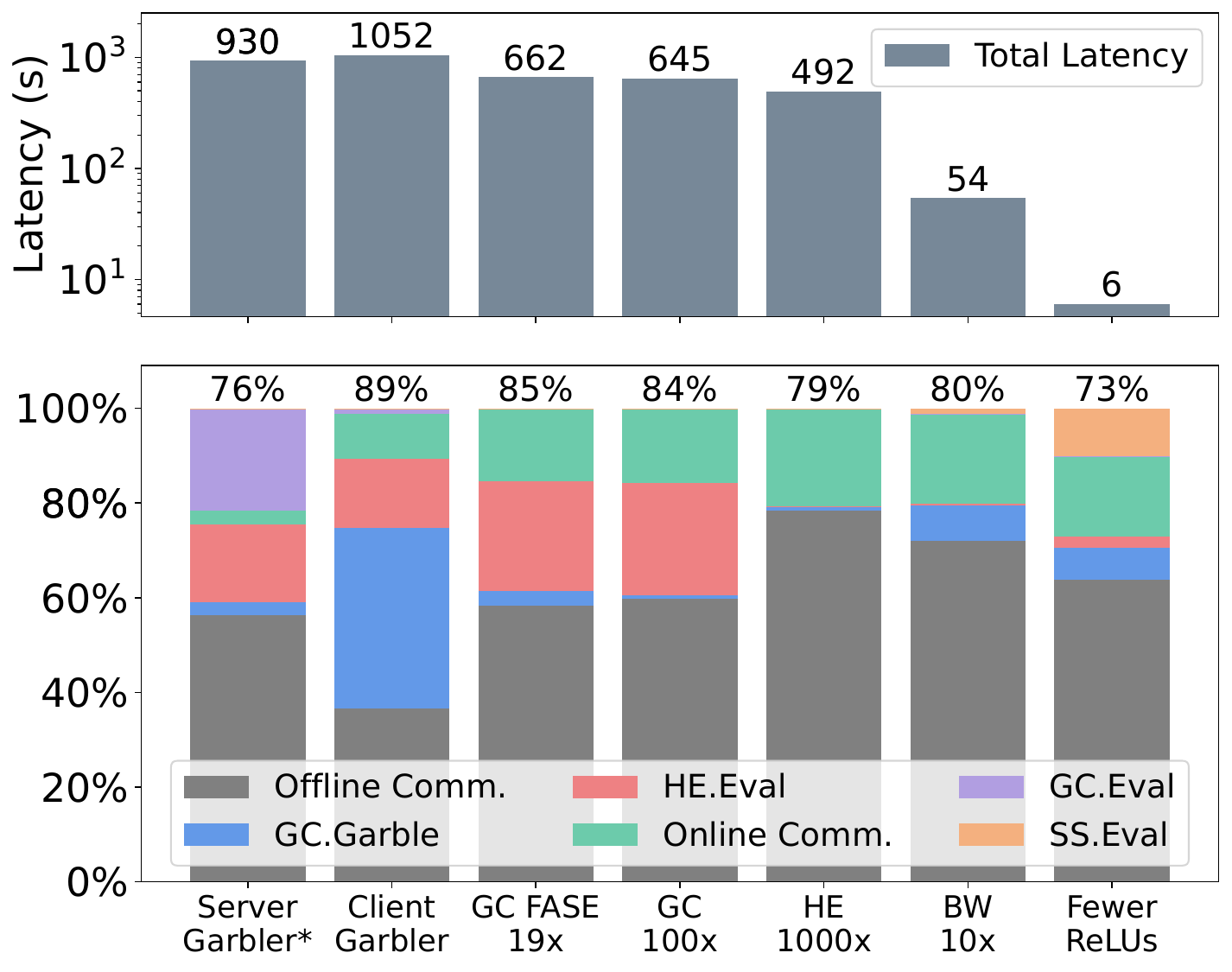} }}
    \vspace{-.5em}
    \caption{Total (offline+online) latency (top) and breakdown of normalized latency (bottom) for different protocols and optimizations. Normalized latency bars are annotated with the percentage of total latency that is offline. 
    * denotes LPHE and WSA are enabled for Server-Garbler.
    }
    \label{fig:analysis}
\end{figure}

In comparing Server-Garbler and Client-Garbler, we first note that the LPHE and WSA optimization are applicable to Server-Garbler as well, 
and reduce its end-to-end latency by 54.6\%.
For a \emph{single} inference, Server-Garbler outperforms Client-Garbler by 13\% in terms of total PI latency, as shown in the first two bars of Figure~\ref{fig:analysis} (top).  
We note that this result is not in contradiction with our findings in the previous sections; with storage constraints and over multiple inference requests, Client-Garbler outperforms Server-Garbler since it is able to better mask offline costs with buffered pre-computes \emph{and} has lower online costs.  In particular, even though Client-Garbler increases online communication latency due to OT  (27.1 seconds to 101 seconds), Client-Garbler shifts the online GC evaluation from the compute-constrained client (modeled as an Intel Atom embedded device) to the powerful server (an AMD EYPC 32-core machine) and reduces GC evaluation from 200 seconds to 11.1 seconds. Thus, the total online latency decreases for Client-Garbler.

How can we further improve the Client-Garbler protocol? From Figure~\ref{fig:analysis} (bottom), we observe that a key bottleneck in Client-Garbler is 
the high overhead of 
circuit garbling on the compute-limited client (the blue component of the Client-Garbler bar).
We now discuss the impact of future research innovations that mitigate this and other performance bottlenecks.

\subsection{Estimating the Benefits of Future Research}
\label{subsec:gc_future}

\textbf{Accelerating GCs:} Our data suggest accelerating garbled circuits would significantly reduce PI computational time for both protocols, assuming we accelerate both the evaluation and garbling phases. 
In the Client-Garbler setting, GC garbling accounts for 38\% of the total PI run-time. 
We first assume the speedups from FASE~\cite{hussain2019fase} (shown in Figure~\ref{fig:analysis} as GC FASE 19$\times$) for both garbling and evaluation. 
We observe a total PI speedup of 1.59$\times$, from 1052 to 662 seconds, largely due to a decrease in client-side garbling as running GC evaluations on the server is relatively fast. 
Furthermore, we assume that a future accelerator provides an additional $\sim 5\times$ speedup over FASE.  
The results are shown in Figure~\ref{fig:analysis} (GC 100$\times$): total PI latency is reduced from 662 to 645 seconds.






\textbf{Accelerating HE:} With the GC computational cost reduced, we see HE (red) emerge as the next major computational bottleneck.
Prior work has been published on accelerators for HE as applied to private inference.
Recent works have shown that special hardware can achieve 3-5 orders of magnitude speedup relative to CPU execution~\cite{reagen2021cheetah, samardzic2021f1, samardzic2022craterlake}.
Since we are working at a high level, we conservatively assume HE can be sped up by 1000$\times$.
HE 1000$\times$ in Figure~\ref{fig:analysis} shows that assuming an HE accelerator further reduces PI latency from 645 seconds to 492 seconds, a speedup of 1.31$\times$.
Here again, the online latency remains largely the same, as HE is processed entirely during the offline phase.
Looking at the remaining time breakdown, we notice that all substantial sources of latency are incurred as communication time due to the MPC protocols.





\textbf{Next Generation Wireless:}
One obvious solution to reduce communication time is to increase bandwidth.
Here we optimistically assume that a future wireless generation, likely well beyond 6G~\cite{asghar2022evolution}, will provide 10$\times$ more bandwidth.
The results are shown in Figure~\ref{fig:analysis} as BW 10$\times$, and we find the end-to-end PI latency is sped up by 9.1$\times$, dropping from 492 to 54 seconds with online-only taking 12 seconds.
Significant bandwidth savings resurface the high GC garbling times, which now take 10\% of the offline latency.
Considering both computational and communication overheads we observe that all significant overheads are now caused by GC ReLUs.




\textbf{PI-Friendly Neural Architectures}
At this point, the performance limitations are due to ReLU.
This is in stark contrast to performance limitations of plaintext inference, where ReLU is often an afterthought. To this end, 
there has been a recent thrust in the ML community to rethink neural architecture design to make them more PI-amenable, namely by reducing the number of ReLUs~\cite{ghodsi2020cryptonas,cho2021sphynx,lou2020safenet,jha2021deepreduce,cho2022selective}.
Significant progress has already been made, and by combining techniques achieving a 10$\times$ ReLU reduction seems well within reach.
In Figure~\ref{fig:analysis}  (Fewer ReLUs), we plot the PI runtime assuming 10$\times$ ReLUs have been successfully removed.
Reducing ReLUs has many positive effects: less garbling and evaluation time, less garbled circuit communication, less OT communication, and less storage pressure.
These benefits culminate in a PI speedup of 9$\times$, bringing PI inference down to 6 seconds, of which only 1.63 seconds is online. And while this is a significant speedup from our starting point, the total latency is still far from achieving plaintext-level speeds. Going forward, we encourage and welcome researchers to find innovative research avenues and design novel approaches to private inference that bridge this gap.







\section{Related Work} \label{sec:RelatedWork}

We now provide a summary of seminal work and recent developments in privacy-preserving computing and PI.

\textbf{HE:}
Since the first successful demonstration of HE in 2009~\cite{gentry2009fully},
many advances have been made.
Today, two classes of HE exist for computing on integers/fixed-point~\cite{fan2012somewhat, brakerski2014leveled, cheon2017homomorphic} and Boolean logic~\cite{ducas2015fhew, chillotti2020tfhe}.
The former is efficient for addition and multiplication whereas the latter can compute arbitrary logic.
Schemes are widely available today thanks to many software implementations~\cite{sealcrypto,halevi2014helib,polyakov2017palisade,cheon2017homomorphic}.



\textbf{GCs:} 
Garbled circuits were first proposed by Yao in 1986~\cite{yao1986generate}.
They have recently received a resurgence of attention thanks to many significant advances in their performance~\cite{beaver1990round, naor1999privacy, pinkas2009secure,kolesnikov2008improved, kolesnikov2014flexor, zahur2015two}.
The optimizations aim to reduce the amount of work per gate by simplifying the computations and total amount of work.
Two of the most widely used optimizations are FreeXOR~\cite{kolesnikov2014flexor} and HalfGate~\cite{zahur2015two}, which are used here.
Like HE, many GC libraries exist~\cite{demmler2015aby, wang2016emp,mohassel2018aby3,ball2019garbled,patra2021aby2}.


\textbf{SS:}
Secret sharing is a general technique first proposed by~\cite{blakley1979safeguarding,shamir1979share}.
It has received significant attention privacy-preserving inference,
including PyTorch-compatible CrypTen~\cite{knott2021crypten}, 
and GPU extension CryptGPU~\cite{tan2021cryptgpu} for secure tensor processing.

{\bf Protocols for PI:} 
Prior work has explored processing inferences using HE only~\cite{gilad2016cryptonets, samardzic2022craterlake}.
These protocols are convenient as privacy primitives are not changed.
However, they typically introduce accuracy loss via the approximation of ReLU, even with complex training~\cite{garimella2021sisyphus} techniques, and cannot leverage LPHE.
To overcome this many have proposed hybrid protocols~\cite{juvekar2018gazelle, liu2017oblivious, mishra2020delphi} that we improve on in this work.
There has also been work on accelerating private computations assuming a trusted-third party (rather than 2PC, as assumed here), which assumes a weaker security model for higher performance~\cite{mohassel2018aby3,chaudhari2019astra,dalskov2019secure,wagh2019securenn,kumar2020cryptflow,patra2020blaze,wagh2021falcon}. Finally, prior work has performed a characterization of MPC protocols for Transformer models \cite{9804616}.



{\bf PI Amenable Networks:}
The ML community has begun exploring ways to design neural networks with fewer ReLUs.
The general approaches taken so far have been to conduct ReLU aware neural architecture searches~\cite{ghodsi2020cryptonas,cho2021sphynx}, prune ReLUs from the networks~\cite{jha2021deepreduce, cho2022selective}, and approximate ReLU computations for cheaper GC implementations~\cite{ghodsi2021circa}. These methods are orthogonal to the work of this paper and can further reduce GC storage and communication costs along with our proposed optimizations.

DELPHI \cite{mishra2020delphi} and AESPA~\cite{aespa} either partly or fully replace ReLUs with polynomial activation functions that are processed using Beaver Triples \cite{beaver1995precomputing}, which are cheaper in both compute and communication than GCs. 
However, recent results show that replacing ReLUs with low-degree polynomials reduces test accuracy, especially for deeper networks \cite{garimella2021sisyphus}. This is also confirmed in Figure 8 by DELPHI \cite{mishra2020delphi}. In this paper, we only consider highly-accurate, ReLU-only deep learning models that are state-of-the-art.

{\bf HE accelerators:}
A variety of hardware accelerators for HW now exist for both FPGAs~\cite{turan2020heaws, roy2019fpga,riazi2020heax} and ASIC~\cite{reagen2021cheetah,samardzic2021f1, samardzic2022craterlake,kim2022bts}.
The design architectures vary and some support fixed or multiple schemes.
However, the conclusions are largely the same: specialized logic is able to provide significant speedup over the CPU, easily achieving the 1000$\times$ assumed in Section~\ref{sec:discussion}. Recently, \cite{van2022client} proposed HE acceleration for resource-constrained clients and massively reduces the PI overheads.




{\bf GC accelerators:}
Prior work has also looked at accelerating GCs.
The solution space includes general, re-programmable accelerators~\cite{songhori2016garbledcpu},
more fixed-logic solutions~\cite{hussain2018maxelerator}, and even application-specific custom GC hardware for multiplication~\cite{hussain2019fase}.
The designs achieve significant speedup by accelerating the core computations of GC gates and leveraging workload parallelism.



\section{Conclusion}

In this paper, we perform an in-depth evaluation of the system-level characteristics for state-of-the-art hybrid PI protocols. 
In contrast to prior work, which considers only individual inferences, we consider arrival rates of inference requests.
Our experiments reveal excessive offline costs that have largely been ignored, including client-storage pressure, high-latency communication, and long-running HE-GC computations.
The characterization using arrival rates reveals these costs cannot remain offline as there is insufficient storage to buffer pre-computations and insufficient time between requests to mask the extreme computation and communication latency.
Leveraging insights from our characterization we propose three novel optimizations to overcome the storage limitations of the client, speed up HE computations, and lessen the communication latency.
In all, our optimizations reduce the total PI time by 1.8$\times$.

We also evaluate the potential of future optimizations across the PI pipeline and their effect on bringing PI close to practicality. 
We hope that our analysis encourages researchers to account for end-to-end system effects in PI optimizations.

\section{Acknowledgements}
We would like to thank Dr. Sundeep Rangan for his advice regarding the imbalance of wireless communication during private inference protocols. We also thank our shepherd and
the anonymous reviewers for their thorough and insightful feedback. This work was supported in part by the NSF under grants 1565396 and 1553419, the Applications Driving Architectures (ADA) Research Center, a JUMP Center co-sponsored by SRC and DARPA. This research was
also developed with funding from the Defense Advanced
Research Projects Agency (DARPA), under the Data Protection in Virtual Environments (DPRIVE) program, contract
HR0011-21-9-0003. The views, opinions and/or findings
expressed are those of the author and should not be interpreted as representing the official views or policies of the
Department of Defense or the U.S. Government.
%
%
%
%
%






\appendix
\section{Artifact Appendix}

\subsection{Abstract}

We open source our private inference simulator at the following GitHub repo: \href{https://github.com/kvgarimella/characterizing-private-inference}{https://github.com/kvgarimella/characterizing-private-inference}. We construct a model of a system for private inference and a simulator using Simpy to explore and evaluate tradeoffs under different system conditions. We model a single-client, single-server setting where inferences are queued in a FIFO manner and are generated by sampling from a Poisson distribution. 

The repository itself contains four high-level directories. The directory \texttt{garbled\_circuits} contains the raw data for benchmarking ReLU Garbling and Evaluation on an Intel Atom Z8350 embedded device (\SI{1.92}{GHz}, 4 cores, \SI{2}{GB} RAM) and an AMD EPYC 7502 server (\SI{2.5}{GHz}, 32 cores, \SI{256}{GB} RAM). These two devices represent our client and server, respectively. Next, the directory \texttt{layer\_parallel\_HE} contains our code and the raw data for applying layer-parallel homomorophic to linear layer evaluations. The directory \texttt{simulator} contains our private inference simulator. Finally, \texttt{artifact} contains scripts to replicate key figures in our paper.

\subsection{Artifact check-list (meta-information)}

{\small
\begin{itemize}

  \item {\bf Program: } Python Simulator for Private Inference 
  \item {\bf Run-time environment: }Linux OS with Python
  \item {\bf Output: }Replication of key figures 
  \item {\bf Experiments: } Compute and storage cost analysis of private inference as well as results for simulating workloads of private inference.
  \item {\bf How much time is needed to prepare workflow (approximately)?: } 15 minutes to clone GitHub repository, install dependencies, and run a simple test.
  \item {\bf How much time is needed to complete experiments (approximately)?: } 3 hours.
  \item {\bf Publicly available?: }\href{https://github.com/kvgarimella/characterizing-private-inference}{https://github.com/kvgarimella/characterizing-private-inference}
  \item {\bf Code licenses: } MIT 
  \item {\bf Archived: \href{https://www.doi.org/10.5281/zenodo.7633678}{https://www.doi.org/10.5281/zenodo.7633678}}
\end{itemize}
}


\subsubsection{How to access} 

The code repository can be accessed at:\\ \href{https://github.com/kvgarimella/characterizing-private-inference}{https://github.com/kvgarimella/characterizing-private-inference}.

\subsubsection{Software dependencies}
Python is required to run the Private Inference simulator. The required Python packages can be found in the \texttt{requirements.txt} file within the GitHub repo. Optionally, Microsoft SEAL, Eigen, and OpenMP are required to benchmark Layer Parallel Homomorphic Encryption within the directory \texttt{layer\_parallel\_HE}.

\subsection{Installation}
First, clone the GitHub repository. Then, follow the instructions in the README to install the required Python packages. As a basic test, navigate to the subdirectory \texttt{simulator/experiments/} and run \texttt{python simulate\_server\_garbler.py}. This should run a single private inference workload experiment and create an output folder named \texttt{tmp} containing results of the simulation.

\subsection{Experiment workflow}
Please refer to the README within the \texttt{artifact} directory. In some cases, simply calling a Python script builds a figure presented in this paper. For other experiments involving simulations, shell scripts are provided to first run the simulations and then parse and plot the results from these simulations.

\subsection{Evaluation and expected results}

The directory \texttt{artifact} contains scripts to reproduce Figures \ref{fig:storage_issue}, \ref{fig:compute_issue}, \ref{fig:non_zero_arrival_rates}, \ref{fig:optimization_storage}, \ref{fig:optimized-compute}, \ref{fig:twostacked}, and \ref{fig:analysis}. Figures \ref{fig:storage_issue} and \ref{fig:compute_issue} show the end-to-end storage and compute overheads of running a single network inference for ResNet-32, VGG-16, and ResNet-18 for the image classification datasets CIFAR, TinyImageNet, and ImageNet. Figure \ref{fig:non_zero_arrival_rates} moves beyond our single inference analysis and introduced workloads of inferences over a period of time. Figures \ref{fig:optimization_storage} and \ref{fig:optimized-compute} show our proposed optimizations for both storage (Client-Garbler) and compute (Layer Parallel Homomorphic Evaluation). Figure \ref{fig:twostacked} compares our optimized protocol to the baseline and is representative of our results for Figures \ref{fig:lphe_vs_rlp} and \ref{fig:sensitivity_sweep}. Finally, Figure \ref{fig:analysis} examines future optimization on end-to-end latency for private inference. 

Navigate to the \texttt{artifact} directory. For each of the figures that can be reproduced, either run the provided python or shell script. Running the scripts simulates private inference workloads, parses the data, and reproduces the plots.

\balance
\bibliographystyle{ACM-Reference-Format}
\bibliography{references}

\end{document}